\documentclass[11pt,a4paper]{article}
\usepackage{times}
\usepackage{a4wide}
\usepackage{amsfonts}
\usepackage{amssymb}
\usepackage{amsmath}
\usepackage{booktabs} % for much better looking tables
\usepackage{array}
\usepackage{ifpdf}
\ifpdf
\usepackage[pdftex,unicode,implicit]{hyperref}
\hypersetup{%
  pdftitle    = {All the timelike supersymmetric solutions
 of gauged \texorpdfstring{$N=2$}{N=2}, \texorpdfstring{$d=4$}{d=4} supergravity},
  pdfkeywords = {supersymmetry, supergravity, special  geometry, BPS solutions},
  pdfauthor   = {Patrick Meessen and  Tomas Ort\'{\i}n},
  plainpages  = true,
  colorlinks  = true,
  citecolor   = blue,
  urlcolor    = red,
  linkcolor   = black
}
\newcommand{\hepth}[1]{{\tt
\href{http://www.arXiv.org/abs/hep-th/#1}{hep-th/#1}}}
\newcommand{\grqc}[1]{{\tt
\href{http://www.arXiv.org/abs/gr-qc/#1}{gr-qc/#1}}}

\newcommand{\arxiv}[1]{{\tt
\href{http://www.arXiv.org/abs/#1}{#1}}}
\else
  \usepackage[dvips]{graphicx}
  \usepackage[unicode,implicit]{hyperref}
  \newcommand{\hepth}[1]{{\tt hep-th/#1}}
  \newcommand{\grqc}[1]{{\tt gr-qc/#1}}
  
  \newcommand{\arxiv}[1]{{\tt arXiv:#1}}
\fi
\usepackage{tikz}
\newcommand{\FPAUO}[2]{
\tikz[scale=.13,
         Uniovi/.style={color=gray, fill=gray}
 ] {
 \fill[Uniovi] (0,0) circle (10);
 \fill[white] (0,7) circle (1.5);
 \draw[Uniovi] (-2,7.5) rectangle (2,5.5);
 \fill[white] (-0.3,6.6) rectangle (0.3,0);   % 1.7 cm 
 \fill[white] ( -0.9,6.2) rectangle (.9 ,5.6);
 \fill[white] (-1.4, 5.2) rectangle (1.4, 4.6);
 \fill[white] (0,0) ellipse (3.5 and 4);
 \fill[Uniovi] (-2.5,0.3) rectangle (2.5,-0.3);
 \fill[Uniovi] (-2,2.3) rectangle (2,1.7);
 \fill[Uniovi] (-2,-2.3) rectangle (2,-1.7);
 \fill[white] (-4.5,5.5) rectangle (-2.7,4.9);
 \fill[white] (-3.9,6.1) rectangle (-3.3,4.3);
 \fill[white] (4.5,5.5) rectangle (2.7,4.9);
 \fill[white] (3.9,6.1) rectangle (3.3,4.3);
 \foreach \x in { 0,..., 3 }
   \foreach \y in { 0,...,\x}
    {
     \fill[white] (-6-\x*0.7+\y*1.4,3.5-\x *1.97) -- (-5.6-\x*0.7+\y*1.4,2.4-\x *1.97) -- (-6.4-\x*0.7+\y*1.4,2.4-\x *1.97) -- cycle;
     \fill[white] (6-\x*0.7+\y*1.4,3.5-\x *1.97) -- (5.6-\x*0.7+\y*1.4,2.4-\x *1.97) -- (6.4-\x*0.7+\y*1.4,2.4-\x *1.97) -- cycle;
   };
 \draw (0,-6) node[
                               text centered, 
                               color=white, 
                               font={\fontsize{8}{4}\sffamily\selectfont}
                             ] {FPAUO-#1/#2};
}} 
\makeatletter
\@addtoreset{equation}{section}
\makeatother

\begin{document}

~\vspace{-4cm}\begin{flushright}
\small
\FPAUO{12}{06}\\
IFT-UAM/CSIC-11-39\\
%\texttt{arXiv:YYMM.NNNN [hep-th]}\\
April 2\textsuperscript{nd}, 2012\\
\normalsize
\end{flushright}

\begin{center}

\vspace{1cm}

%Original title: All the timelike supersymmetric solutions of gauged
%N=2, d=4 supergravity
{\LARGE {\bf Supersymmetric solutions to gauged $N=2$ $d=4$ SUGRA:}}\\[.5cm] 
{\LARGE {\bf the full timelike shebang}}

\vspace{2cm}

 {\sl\large Patrick Meessen}
 \footnote{E-mail: {\tt meessenpatrick  [at] uniovi.es}}$^{\dagger}$,
{\sl\large and Tom{\'a}s Ort\'{\i}n}
\footnote{E-mail: {\tt Tomas.Ortin  [at] csic.es}}$^{\ddagger}$

\vspace{.8cm}

${}^{\dagger}${\it HEP Theory Group, Departamento de F\'{\i}sica, Universidad de Oviedo\\ 
        Avda.~Calvo Sotelo s/n, 33007 Oviedo, Spain}\\

\vspace{.3cm}

$^{\ddagger}${\it Instituto de F\'{\i}sica Te\'orica UAM/CSIC\\
C/ Nicol\'as Cabrera, 13-15,  C.U.~Cantoblanco,  E-28049-Madrid, Spain}

\vspace{2cm}

%%%%%%%%%%%%%%%%%%%%%%%%%%%%%%%%%%%%%%%%%%%%%%%%%%%%%%%%%%%%%%%%%%%%%%

{\bf Abstract}

\end{center}

\begin{quotation}\small
  We discuss the structure of supersymmetric solutions in the timelike case to
  general gauged $N=2$ $d=4$ supergravity theories coupled to non-Abelian
  vector multiplets and hypermultiplets.
\end{quotation}

\newpage
%%%%%%%%%%%%%%%%%%%%%%%%%%%%%%%%%%%%%%%%%%%%%%%%%%%%%%%%%%%%%%%%%%%%%%
%%%%%%%%%%%%%%%%%%%%%%%%%%%%%%%%%%%%%%%%%%%%%%%%%%%%%%%%%%%%%%%%%%%%%%
%%%%%%%%%%%%%%%%%%%%%%%%%%%%%%%%%%%%%%%%%%%%%%%%%%%%%%%%%%%%%%%%%%%%%%
%%%%%%%%%%%%%%%%%%%%%%%%%%%%%%%%%%%%%%%%%%%%%%%%%%%%%%%%%%%%%%%%%%%%%%
\pagestyle{plain}
%%%%%%%%%%%%%%%%%%%%%%%%%%%%%%%%%%%%%%%%%%%%%%%%%%%%%%%%%%%%%%%%%%%%%%
%%%%%%%%%%%%%%%%%%%%%%%%%%%%%%%%%%%%%%%%%%%%%%%%%%%%%%%%%%%%%%%%%%%%%%
%%%%%%%%%%%%%%%%%%%%%%%%%%%%%%%%%%%%%%%%%%%%%%%%%%%%%%%%%%%%%%%%%%%%%%
%%%%%%%%%%%%%%%%%%%%%%%%%%%%%%%%%%%%%%%%%%%%%%%%%%%%%%%%%%%%%%%%%%%%%%
%%%%%%%%%%%%%%%%%%%%%%%%%%%%%%%%%%%%%%%%%%%%%%%%%%%%%%%%%%%%%%%%%%%%%%

\tableofcontents

%%%%%%%%%%%%%%%%%%%%%%%%%%%%%%%%%%%%%%%%%%%%%%%%%%%%%%%%%%%%%%%%%%%%%%
%%%%%%%%%%%%%%%%%%%%%%%%%%%%%%%%%%%%%%%%%%%%%%%%%%%%%%%%%%%%%%%%%%%%%%
%%%%%%%%%%%%%%%%%%%%%%%%%%%%%%%%%%%%%%%%%%%%%%%%%%%%%%%%%%%%%%%%%%%%%%
%%%%%%%%%%%%%%%%%%%%%%%%%%%%%%%%%%%%%%%%%%%%%%%%%%%%%%%%%%%%%%%%%%%%%%
\section*{Introduction}
%%%%%%%
The starting shot for the systematic characterization of supersymmetric
solutions to supergravity theories was given in 1982 by Gibbons {\&} Hull
\cite{Gibbons:1982fy}, who obtained a partial characterization of the
supersymmetric solutions of pure (minimal) $N=2$ $d=4$ supergravity, later
completed by Tod \cite{Tod:1983pm} using the consistency conditions of the
Killing spinor equations, who realized that the assumption of hypesurface
orthogonality implicitly made by Gibbons and Hull was unnecessary for a
solution to be supersymmetric.  In a related development, in 1985
Kowalski-Glikman \cite{KowalskiGlikman:1985im} proved that the only solutions
to minimal $N=2$ $d=4$ SUGRA that do not break any supersymmetries, called the
maximally supersymmetric solutions, are Minkowski space, the Robinson-Bertotti
spacetime ($aDS_{2}\times S^{2}$) and a specific pp-wave called the
4-dimensional Kowalski-Glikman wave.
\par
In the 30 years since, hosts of results\footnote{ We feel that it is sheer
  impossible to give an overview doing justice to all the interesting results
  obtained and shall restrict ourselves mainly to the results concerning the
  classification of supersymmetric solutions to $N=2$ $d=4$ supergravity and
  apologize in advance for any omission.  }  concerning supersymmetric
solutions to SUGRAs have been obtained and many potent techniques were
developed in order to obtain them. The first of such techniques was developed
by Gauntlett {\em et al.\/} in ref.~\cite{Gauntlett:2002nw} and used it to
give a complete classification of supersymmetric solutions to minimal $N=1$
$d=5$ SUGRA; this technique goes by the name of {\em bilinear method} as it
deals with the analysis of all the form-fields one can construct as bilinears
out off the Killing spinors. In this method there are 2 types of relations for
the bilinears: first of all there are ``kinematical'' relations between
products of bilinears due to the Fierz identities and only depend on the
number and type of spinors employed in a given theory, and not on the theory
itself (matter content, equations of motion etc.). The second kind of
relations are ``dynamical'' in that they are differential relations
determining the spacetime dependency of the bilinears and which originate in
the theory-specific Killing spinor equations.  In this article we will use the
bilinear method, but it must be mentioned that there are more techniques {\em
  e.g.\/} {\em spinorial geometry} proposed by Gillard {\em et al.\/} in
ref.~\cite{Gillard:2004xq}, which is extremely powerful and much of the
progress in the characterization of supersymmetric solutions, especially to
higher dimensional SUGRAs, was made using it.
\par
Another technique that was developed in ref.~\cite{Gauntlett:2002nw} is an
effort-saving one: preserved supersymmetry implies, by means of the
integrability condition for the existence of a Killing spinor, relations
between the equations motions implying that there is a minimal set of
(components of) equations of motion that, once these are satisfied,
automatically ensure that all the equations of motion are satisfied.  In
ref.~\cite{Bellorin:2005hy} this effort-saving technique was linked to the
so-called {\em Killing Spinor Identities} originally derived in
ref.~\cite{Kallosh:1993wx}; the KSIs are the restriction of the gauge
identities expressing the fact that a SUGRA action is supersymmetric, to the
case of vanishing fermionic fields and with a gauge parameter taken to be the
Killing spinor. The bottom line of \cite{Bellorin:2005hy}'s identification is
that there is no need to calculate the integrability conditions and that only
the supersymmetric variations of the bosonic fields need to be known.
\par
The generalization of Gibbons {\&} Hull's result to vector multiplet-coupled
$N=2$, $d=4$ SUGRA was done first for static spacetimes in
refs.~\cite{Sabra:1997kq} and in ref.~\cite{Behrndt:1997ny} for general
stationary solutions. In ref.~\cite{Meessen:2006tu} the authors carried out a
full characterization of supersymmetric solutions to this theory finding in
the timelike case full agreement with aforementioned works; the null case was
found to allow not only for pp-waves but also for stringy cosmic strings of
the type first studied in \cite{Greene:1989ya}.  This characterization was
then extended to the case of $N=2$, $d=4$ SUGRA coupled to vector multiplets
and hypermultiplets in ref.~\cite{Huebscher:2006mr} and to the case of
YM-vector multiplets in refs.~\cite{Huebscher:2007hj,Huebscher:2008yz}; in the
latter theories one can, depending on the model, construct analytic, globally
regular monopole solutions and non-Abelian black holes.  Caldarelli {\&} Klemm
\cite{Caldarelli:2003pb} extended Tod's results to the case of minimal gauged
$N=2$, $d=4$ supergravity and the resulting solutions were studied further by
Cacciatori {\em et al.\/} in refs.~\cite{Cacciatori:2004rt}; some examples of
supersymmetric black holes had been already been obtained in
refs.~\cite{Romans:1991nq,Sabra:1999ux}.  The fact that the maximally
supersymmetric solution to this theory is $aDS_{4}$, was established by
Kowalski-Glikman in ref.~\cite{KowalskiGlikman:1985wi} and in
refs.~\cite{Grover:2006wy} it was shown that all solutions preserving more
than half of the supersymmetry necessarily arise as quotients of $aDS_{4}$.
In refs.~\cite{Cacciatori:2008ek} the characterization of supersymmetric
solutions to minimal gauged $N=2$, $d=4$ SUGRA was extended by considering the
coupling to Abelian vector multiplets and (rotating) black hole solutions were
constructed in refs.~\cite{Cacciatori:2009iz}.  Finally, in
refs.~\cite{Meessen:2009ma} a classification was made for the fake-SUGRA
analogue of gauged minimal $N=2$, $d=4$ SUGRA coupled to YM-vector multiplets,
leading to generalizations of Kastor {\&} Traschen's cosmological multi-black
hole solutions \cite{Kastor:1992nn}; as these theories are obtained by
Wick-rotating the $\mathrm{U}(1)$ Fayet-Iliopoulos term, the potential has the
opposite sign w.r.t.~supersymmetric theory, the maximally fake-supersymmetric
solution is 4-dimensional De Sitter space.
\par
What for the moment is missing from the above laundry list of classification
articles is the characterization of supersymmetric solutions to gauged $N=2$,
$d=4$ supergravity coupled to YM-vector multiplets and hypermultiplets: the
aim of the current article is to do just that, albeit for the timelike case
only.
\par
Observe that this is no way means that there are no supersymmetric solutions
to the full theory known to the literature: for example supersymmetric domain
walls were constructed in ref.~\cite{Behrndt:2001mx}, recently the maximally
supersymmetric solutions were classified in ref.~\cite{Hristov:2009uj} and
supersymmetric Lifschitz, Schr\"odinger and (anti-)De Sitter solutions were
considered in refs.~\cite{Halmagyi:2011xh,Kachru:2011ps}; Supersymmetric
black-hole solutions with an Abelian gauging were constructed in
ref.~\cite{Hristov:2010eu} and further analyzed in ref.~\cite{Hristov:2010ri}.
Lastly, let us mention refs.~\cite{Hristov:2011ye} in which the Bogomol'nyi
bound for asymptotically $aDS_{4}$ black holes and black strings are
discussed.
\par
The outline of this article follows the algorithm used in the classification
of supersymmetric solutions and is as follows: section \ref{sec-thetheory}
contains an extremely short introduction to gauged $N=2$, $d=4$ SUGRA coupled
to YM-vector and hypermultiplets. In section \ref{sec:KSIs} we formulate the
basic problem of finding SUGRA solutions preserving some supersymmetry as the
problem of finding expressions for the purely bosonic SUGRA fields that allow
for supersymmetry transformations that do not generate non-trivial fermionic
fields; the relevant equations to be solved are called the Killing Spinor
Equations (KSEs) and the supersymmetric variation parameter is called the
Killing Spinor.  Given that information, we detail the KSIs and discuss the
minimal set of (components of) equations that must be checked explicitly as to
be sure that all the equations of motion are satisfied.  Then in
section~\ref{sec-bilinearkse} we analyze the differential constraints on the
bilinears\footnote{ As was mentioned above, the implications of the Fierz
  identities do not depend on the matter couplings and we shall take them as
  given and refer the reader to ref.~\cite{Bellorin:2005zc} for more
  information.} in general and in section~\ref{sec-timelikemetric} we shall
introduce coordinates and obtain the restrictions on the metric; at the end of
that section we shall have obtained necessary conditions on the fields in our
theory for them to give non-trivial solutions to the KSEs.  In
section~\ref{sec:suffi} we will show that the conditions obtained thus far are
not only necessary but also sufficient to guarantee preserved supersymmetry.
In section~\ref{sec:SusySol} we shall then discuss the equations that need to
be satisfied in order to solve the SUGRA equations of motion.  {}Finally,
section \ref{sec-conclusions} contains our conclusions.
\par
We could not resist the temptation to include some appendices which explain
the meaning and properties of the mathematical objects which are used in the
gauging of $N=2$, $d=4$ theories: in app.~\ref{sec-sigmas} we give
relations for the Pauli matrices and how to decompose the various spinorial
bilinears using them.  App.~\ref{app-SKGgauging} deals with the gauging of
isometries in Special Geometry and app.~\ref{app-QMgauging} does the same but
for the hypermultiplets.
%%%%%%%%%%%%%%%%%%%%%%%%%%%%%%%%%%%%%%%%%%%%%%%%%%%%%%%%%%%%%%%%%%%%%%
%%%%%%%%%%%%%%%%%%%%%%%%%%%%%%%%%%%%%%%%%%%%%%%%%%%%%%%%%%%%%%%%%%%%%%
%%%%%%%%%%%%%%%%%%%%%%%%%%%%%%%%%%%%%%%%%%%%%%%%%%%%%%%%%%%%%%%%%%%%%%
%%%%%%%%%%%%%%%%%%%%%%%%%%%%%%%%%%%%%%%%%%%%%%%%%%%%%%%%%%%%%%%%%%%%%%
%%%%%%%%%%%%%%%%%%%%%%%%%%%%%%%%%%%%%%%%%%%%%%%%%%%%%%%%%%%%%%%%%%%%%%
\section{General gauged $N=2$, $d=4$ supergravity }
\label{sec-thetheory}
%%%%
In this section we are going to give a brief description of $N=2$, $d=4$
supergravity coupled to $n$ vector supermultiplets and $m$ hypermultiplets
with gaugings of some of the isometries of the scalar manifolds associated to
perturbative symmetries of the whole theory\footnote{
  See {\em e.g.\/}
  ref.~\cite{Andrianopoli:1996cm}, the review \cite{kn:toinereview} and the
  original works \cite{deWit:1984pk,deWit:1984px} for more information on
  $N=2$, $d=4$ supergravities. Our conventions are explained in
  refs.~\cite{Bellorin:2005zc,Meessen:2006tu,Huebscher:2006mr,Huebscher:2008yz}.
}
using as gauge fields the fundamental (electric) vectors.\footnote{
  The
  embedding-tensor formalism introduced in refs.~\cite{Cordaro:1998tx}
  offers more general possibilities. Some of them have been explored in
  refs.~\cite{de Vroome:2007zd}.
}
 
The gravity multiplet of the $N=2$, $d=4$ theory consists of the graviton
$e^{a}{}_{\mu}$, a pair of gravitinos $\psi_{I\, \mu}$ ($I=1,2$), which
we describe as Weyl spinors, and a vector field $A^{0}{}_{\mu}$ (the
\textit{graviphoton}).

Each of the $n$ vector supermultiplets of $N=2$, $d=4$ supergravity, labeled by
$i,j,k=1,\cdots, n$ contains one complex scalar $Z^{i}$, a pair of gaugini
$\lambda^{I\, i}$ described as Weyl spinors, and a vector field
$A^{i}{}_{\mu}$. The $\bar{n}=n+1$ vectors $A^{0}{}_{\mu},A^{i}{}_{\mu}$ are
are described collectively by an array $A^{\Lambda}{}_{\mu}$
($\Lambda=0,\cdots,n$). In the ungauged theory, the scalar
self-coupling
is described by a non-linear $\sigma$-model with K\"ahler metric
$\mathcal{G}_{ij^{*}}(Z,Z^{*})$; their coupling to the vector fields
by means of a
complex matrix $\mathcal{N}_{\Lambda\Sigma}(Z,Z^{*})$.  These two couplings
are related by a structure called Special K\"ahler Geometry.\footnote{
  See {\em e.g.\/} ref.~\cite{Andrianopoli:1996cm} or the appendix of
  ref.~\cite{Meessen:2006tu}.
} 
In the gauged theory there are additional
couplings due to the scalar potential and the covariant derivatives of the scalars
that depend on the holomorphic components of the Killing vectors
$k_{\Lambda}{}^{i}(Z)$ generating the isometries that have been gauged, and
on the momentum map $\mathcal{P}_{\Lambda}(Z,Z^{*})$, defined in
eq.~(\ref{eq:SKGmomentummapdef}). The gauging of the isometries of the Special
K\"ahler Geometry are described in detail in Appendix~\ref{app-SKGgauging}.

Each hypermultiplet consists of 4 real scalars $q$ (called \textit{hyperscalars}) and
2 Weyl spinors $\zeta$ called \textit{hyperini}. The $4m$ hyperscalars are
collectively denoted by $q^{u}$ ($u=1,\ldots,4m$) and the $2m$ hyperini
by $\zeta_{\alpha}$ ($\alpha =1,\ldots,2m$).  The $4m$ hyperscalars
parametrize a Quaternionic K\"ahler manifold with metric
$\mathsf{H}_{uv}(q)$. In the ungauged theory, the hyperscalars do not couple
directly to any of the fields of the vector multiplets and their only
self-coupling is determined by $\mathsf{H}_{uv}(q)$. In the gauged theory,
however, there are direct couplings between the hyperscalars and the vectors and
complex scalars in the scalar potential and in the covariant derivatives of
the hyperscalars. These couplings depend on the the Killing vectors
$\mathsf{k}_{\Lambda}{}^{u}(q)$ of the isometries that have been gauged and on
the triholomorphic momentum map $\mathsf{P}_{\Lambda}{}^{x}(q)$, defined in
eq.~(\ref{eq:triholomomentummapdef}). The gauging of the isometries of the
Quaternionic K\"ahler manifold are described in
Appendix~\ref{app-QMgauging}.

The action of the bosonic fields of the theory is \cite{Andrianopoli:1996cm}
\begin{equation}
\label{eq:actionhyper}
\begin{array}{rcl}
  S & = & {\displaystyle\int} d^{4}x \sqrt{|g|}
  \left[R 
    +2\mathcal{G}_{ij^{*}}\mathfrak{D}_{\mu}Z^{i}\mathfrak{D}^{\mu}Z^{*\, j^{*}}
    +2 \mathsf{H}_{uv} \mathfrak{D}_{\mu} q^{u} \mathfrak{D}^{\mu} q^{v} 
    +2\Im{\rm m}\mathcal{N}_{\Lambda\Sigma} 
    F^{\Lambda\, \mu\nu}F^{\Sigma}{}_{\mu\nu}
  \right. \\
  & & \\
  & & \left. 
    \hspace{2cm}
    -2\Re{\rm e}\mathcal{N}_{\Lambda\Sigma}  
    F^{\Lambda\, \mu\nu}\star F^{\Sigma}{}_{\mu\nu}
    -V(Z,Z^{*},q)
  \right]\, ,
\end{array}
\end{equation}
where the covariant derivatives acting on the scalars are defined in
eqs.~(\ref{eq:nablazio}) and~(\ref{eq:Dq}), the vector field strengths in
eq.~(\ref{eq:Fdef}); the scalar potential $V(Z,Z^{*},q)$ is given by
\begin{equation}
\label{eq:potgen2}
\begin{array}{rcl}
  V(Z,Z^{*},q) & = & 
g^{2}\left[
-{\textstyle\frac{1}{4}} 
(\Im{\rm m}\mathcal{N})^{-1|\Lambda\Sigma}\mathcal{P}_{\Lambda}\mathcal{P}_{\Sigma}
+{\textstyle\frac{1}{2}} \mathcal{L}^{*\,\Lambda} \mathcal{L}^{\Sigma}
(4\mathsf{H}_{uv}\mathsf{k}_{\Lambda}{}^{u} \mathsf{k}_{\Sigma}{}^{v}
-3\mathsf{P}_{\Lambda}{}^{x} \mathsf{P}_{\Sigma}{}^{x})
\right.
\\
& & \\
& & 
\hspace{.5cm}\left.
+{\textstyle\frac{1}{2}}\mathcal{G}^{ij^{*}}f^{\Lambda}{}_{i}f^{*\, \Sigma}{}_{j^*}
\mathsf{P}_{\Lambda}{}^{x} \mathsf{P}_{\Sigma}{}^{x}\ \right]\, .
\end{array}
\end{equation}

The supersymmetry transformation rules of the fermions for vanishing fermions
are

\begin{eqnarray}
\label{eq:gravisusyrulehyper}
\delta_{\epsilon}\psi_{I\, \mu} 
& = & 
\mathfrak{D}_{\mu}\epsilon_{I} 
+
\left[
T^{+}{}_{\mu\nu}\varepsilon_{IJ}
-\tfrac{1}{2}S^{x}\eta_{\mu\nu}\varepsilon_{IK}(\sigma^{x})^{K}{}_{J}
\right]
\gamma^{\nu}\epsilon^{J}\, ,
\\
& & \nonumber \\
\label{eq:gaugsusyrulehyper}
\delta_{\epsilon}\lambda^{Ii} 
& = & 
i\not\!\!\mathfrak{D} Z^{i}\epsilon^{I} 
+
\left[
\left(\not\!G^{i\, +} +W^{i}\right)\varepsilon^{IJ}
+\tfrac{i}{2}W^{i\, x}\, (\sigma^{x})^{I}{}_{K}\varepsilon^{KJ} 
\right]
\epsilon_{J}\, ,\\
& & \nonumber \\
\label{eq:hypersusyrulehyper}
\delta_{\epsilon}\zeta_{\alpha} & = & 
i\mathsf{U}_{\alpha I\, u}\not\!\!\mathfrak{D}q^{u}\epsilon^{I}
+N_{\alpha}{}^{I}\epsilon_{I}\, ,
\end{eqnarray}

\noindent
where the covariant derivative acting on the spinors is given in
eq.~(\ref{eq:covariantderivativeonepsilon}), our conventions for the Pauli
matrices are described in Appendix~\ref{sec-sigmas} and where $T_{\mu\nu}$ and
$G^{i}{}_{\mu\nu}$ are, respectively, the graviphoton and matter vector field
strengths; they are defined by
\begin{align}
T_{\mu\nu} 
& \equiv 
2i\mathcal{L}^{\Sigma}\Im{\rm m}\, 
\mathcal{N}_{\Sigma\Lambda}F^{\Lambda}{}_{\mu\nu}\, ,
\\
& \nonumber \\
G^{i\, +}{}_{\mu\nu} 
& \equiv
-\mathcal{G}^{ij^{*}}
f^{*\, \Sigma}{}_{j^{*}}\Im{\rm m}\, 
\mathcal{N}_{\Sigma\Lambda}F^{\Lambda}{}_{\mu\nu}
\, .
\end{align}
%
%\noindent
Th so-called \textit{fermion shifts} $S^{x},W^{i},W^{i\ x},N_{\alpha}{}^{I}$
are given by

\begin{eqnarray}
\label{eq:Sx}
S^{x} & = & {\textstyle\frac{1}{2}} g
\mathcal{L}^{\Lambda} \mathsf{P}_{\Lambda}{}^{x}\, ,
\\
& & \nonumber\\
\label{eq:Wi}
W^{i} & = & 
{\textstyle\frac{1}{2}}g\mathcal{L}^{*\, \Lambda}k_{\Lambda}{}^{i}
=
-{\textstyle\frac{i}{2}}
g\mathcal{G}^{ij^{*}}f^{*\Lambda}{}_{j^{*}} \mathcal{P}_{\Lambda}\, ,
\\
& & \nonumber\\
\label{eq:Wix}
W^{i\,x} & = &  g\mathcal{G}^{ij^{*}} f^{*\,\Lambda}{}_{j^{*}}
\mathsf{P}_{\Lambda}{}^{x}\, ,
\\
& & \nonumber\\
\label{eq:NaI}
N_{\alpha}{}^{I} & = & g\mathsf{U}_{\alpha}{}^{I}{}_{u}\mathcal{L}^{*\,\Lambda} 
\mathsf{k}_{\Lambda}{}^{u}\, .
\end{eqnarray}

The supersymmetry transformations of the bosonic fields, also for vanishing
fermions, are

\begin{eqnarray}
\label{eq:susytranse}
\delta_{\epsilon} e^{a}{}_{\mu} & = & 
-{\textstyle\frac{i}{4}} \bar{\psi}_{I\, \mu}\gamma^{a}\epsilon^{I}
+\mathrm{c.c.}\, ,\\
& & \nonumber \\ 
\label{eq:susytransA}
\delta_{\epsilon} A^{\Lambda}{}_{\mu} & = & 
{\textstyle\frac{1}{4}}
\mathcal{L}^{\Lambda\, *}
\varepsilon^{IJ}\bar{\psi}_{I\, \mu}\epsilon_{J}
+
{\textstyle\frac{i}{8}}f^{\Lambda}{}_{i}\varepsilon_{IJ}
\bar{\lambda}^{Ii}\gamma_{\mu}
\epsilon^{J}
+\mathrm{c.c.}\, ,\\
& & \nonumber \\
\label{eq:susytransZ}
  \delta_{\epsilon} Z^{i} & = & 
{\textstyle\frac{1}{4}} \bar{\lambda}^{Ii}\epsilon_{I}\, ,\\
& & \nonumber \\
\label{eq:susytransq}
\delta_{\epsilon}q^{u} & = & 
{\textstyle\frac{1}{4}} \mathsf{U}_{\alpha I}{}^{u}
\overline{\zeta}^{\alpha}\epsilon^{I}
+\mathrm{c.c.}\, , 
\end{eqnarray}

\noindent
and do not depend on the gauge coupling constant $g$. Actually, they take the
same form in the gauged and ungauged cases, a fact that will be exploited in
the derivation of the KSIs.

For convenience, we denote the bosonic equations of motion by

\begin{equation}
\mathcal{E}_{a}{}^{\mu}
\equiv 
-\tfrac{1}{2\sqrt{|g|}}\frac{\delta S}{\delta e^{a}{}_{\mu}}\, ,
\hspace{.4cm}
\mathcal{E}_{i} 
\equiv 
-\tfrac{1}{2\sqrt{|g|}}
\frac{\delta S}{\delta Z^{i}}\, ,
\hspace{.4cm}
\mathcal{E}_{\Lambda}{}^{\mu}
\equiv 
\tfrac{1}{8\sqrt{|g|}}\frac{\delta S}{\delta A^{\Lambda}{}_{\mu}}\, ,
\hspace{.4cm}
\mathcal{E}^{u}
\equiv 
-\tfrac{1}{4\sqrt{|g|}}\mathsf{H}^{uv}\frac{\delta S}{\delta q^{v}}\, .
\end{equation}

\noindent
and the Bianchi identities for the vector field strengths by

\begin{equation}
\label{eq:BL}
\mathcal{B}^{\Lambda\, \mu} \equiv \mathfrak{D}_{\nu} \star F^{\Lambda\,
  \nu\mu}\, .  
\end{equation}

\noindent
Using the action eq.~(\ref{eq:actionhyper}), we can calculate them to be
of the form

\begin{eqnarray}
\mathcal{E}_{\mu\nu} 
& = & 
G_{\mu\nu}
+8\Im {\rm m}\mathcal{N}_{\Lambda\Sigma}
F^{\Lambda\, +}{}_{\mu}{}^{\rho}F^{\Sigma\, -}{}_{\nu\rho}
+2\mathcal{G}_{ij^{*}}[\mathfrak{D}_{(\mu}Z^{i} \mathfrak{D}_{\nu )}Z^{*\, j^{*}}
-{\textstyle\frac{1}{2}}g_{\mu\nu}
\mathfrak{D}_{\rho}Z^{i}\mathfrak{D}^{\rho}Z^{*\, j^{*}}]
\nonumber \\
& & \nonumber \\
& & 
+ 2\mathsf{H}_{uv}\ [\mathfrak{D}_{\mu}q^{u}\mathfrak{D}_{\nu}q^{v} 
-{\textstyle\frac{1}{2}}g_{\mu\nu}\mathfrak{D}_{\rho}q^{u}\mathfrak{D}_{\rho}q^{v}]
+{\textstyle\frac{1}{2}}g_{\mu\nu}V(Z,Z^{*},q)\, ,
\label{eq:Emnn2}\\
& & \nonumber \\
\mathcal{E}_{\Lambda}{}^{\mu} 
& = &
\mathfrak{D}_{\nu} \star F_{\Lambda}{}^{\nu\mu}
+{\textstyle\frac{1}{4}}g
(k_{\Lambda\, i^{*}}\mathfrak{D}^{\mu} Z^{*}{}^{i^{*}}+
 k^{*}_{\Lambda\, i}\mathfrak{D}^{\mu} Z^{i} )
+{\textstyle\frac{1}{2}}g
\mathsf{k}_{\Lambda\, u}\mathfrak{D}^{\mu} q^{u}\, ,
\label{eq:EmL}\\
& & \nonumber \\
\mathcal{E}^{i} 
& = & 
\mathfrak{D}^{2}Z^{i} 
+\partial^{i}
F_{\Lambda}{}^{\mu\nu}\star F^{\Lambda}{}_{\mu\nu}
+{\textstyle\frac{1}{2}} \partial^{i}V(Z,Z^{*},q)\, .
\label{eq:Ei2} 
\\
& & \nonumber \\
\label{eq:Eu}
\mathcal{E}^{u} 
& = & 
\mathfrak{D}^{2}q^{u} +{\textstyle\frac{1}{4}}\partial^{u}V(Z,Z^{*},q)\, ,  
\end{eqnarray}

\noindent
where we have defined the dual field strengths

\begin{equation}
F_{\Lambda\,  \mu\nu} \equiv -\tfrac{1}{4\sqrt{|g|}} 
\frac{\delta S}{\delta \star F^{\Lambda}{}_{\mu\nu}}
=
2\Re{\rm e}\, (\mathcal{N}^{*}_{\Lambda\Sigma}F^{\Sigma\, +}{}_{\mu\nu})
= 
\Re{\rm e}\, \mathcal{N}_{\Lambda\Sigma}F^{\Sigma}{}_{\mu\nu}
+
\Im{\rm m}\mathcal{N}_{\Lambda\Sigma}\, \star F^{\Sigma}{}_{\mu\nu}\, .  
\end{equation}

Combining the fundamental vector field strengths $F^{\Lambda}$ with their magnetic duals 
$F_{\Lambda}$ into a symplectic vector 
$\mathcal{F}^{T}=\left( F^{\Lambda},F_{\Lambda}\right)$,
% $\mathcal{F} \equiv \left(
%   \begin{array}{c}
%     F^{\Lambda} \\ F_{\Lambda}\\
%   \end{array}
% \right)$ 
one can rewrite many objects in a manifestly symplectic-invariant
form. For instance, the graviphoton and matter field strengths are given by
\begin{equation}
  \label{eq:2}
  T^{+}{}_{\mu\nu} \; =\; \langle\, \mathcal{V} \mid
  \mathcal{F}_{\mu\nu} \, \rangle
  \hspace{.6cm}\mbox{and}\hspace{.6cm}
  G^{i\, +}{}_{\mu\nu} \; =\; 
  {\textstyle\frac{i}{2}}
\mathcal{G}^{ij^{*}}
\langle\, \mathcal{D}_{j^{*}}\mathcal{V}^{*} \mid \mathcal{F}_{\mu\nu}\, \rangle\, ,
\end{equation}
%
% \begin{align}
% T^{+}{}_{\mu\nu} 
% & \equiv 
% \langle\, \mathcal{V} \mid \mathcal{F}_{\mu\nu} \, \rangle\, ,
% \\
% & \nonumber \\
% G^{i\, +}{}_{\mu\nu} 
% & \equiv
% {\textstyle\frac{i}{2}}
% \mathcal{G}^{ij^{*}}
% \langle\, \mathcal{D}_{j^{*}}\mathcal{V}^{*} \mid \mathcal{F}_{\mu\nu}\, \rangle\, ,
% \end{align}
%
where the symplectic notation is for example explained in ref.~\cite{Andrianopoli:1996cm}.

%%%%%%%%%%%%%%%%%%%%%%%%%%%%%%%%%%%%%%%%%%%%%%%%%%%%%%%%%%%%%%%%%%%%%%
%%%%%%%%%%%%%%%%%%%%%%%%%%%%%%%%%%%%%%%%%%%%%%%%%%%%%%%%%%%%%%%%%%%%%%
%%%%%%%%%%%%%%%%%%%%%%%%%%%%%%%%%%%%%%%%%%%%%%%%%%%%%%%%%%%%%%%%%%%%%%
%%%%%%%%%%%%%%%%%%%%%%%%%%%%%%%%%%%%%%%%%%%%%%%%%%%%%%%%%%%%%%%%%%%%%%
\section{Supersymmetric configurations and Killing Spinor Identities}
\label{sec:KSIs}
%%%%%%
Our first goal is to find all the bosonic field configurations
\begin{displaymath}
\left\{g_{\mu\nu}(x),F^{\Lambda}{}_{\mu\nu}(x),
  A^{\Lambda}{}_{\mu}(x),Z^{i}(x),q^{u}(x)\right\}\, ,
\end{displaymath}

\noindent
for which the Killing Spinor Equations (KSEs) of these theories, {\em i.e.\/}
\begin{eqnarray}
\label{eq:KSE1N2}
\delta_{\epsilon}\psi_{I\, \mu} & = & 
\mathfrak{D}_{\mu}\epsilon_{I} 
+
\left[
T^{+}{}_{\mu\nu}\varepsilon_{IJ}
-\tfrac{1}{2}S^{x}\eta_{\mu\nu}\varepsilon_{IK}(\sigma^{x})^{K}{}_{J}
\right]
\gamma^{\nu}\epsilon^{J}
=0\, ,
\\
& & \nonumber \\
\label{eq:KSE2N2}
\delta_{\epsilon}\lambda^{Ii} & = & 
i\not\!\!\mathfrak{D} Z^{i}\epsilon^{I} 
+
\left[
\left(\not\!G^{i\, +} +W^{i}\right)\varepsilon^{IJ}
+\tfrac{i}{2}W^{i\, x}\, (\sigma^{x})^{I}{}_{K}\varepsilon^{KJ} 
\right]
\epsilon_{J}
=0\, ,
\\
& & \nonumber \\
\label{eq:KSE3N2}
\delta_{\epsilon}\zeta_{\alpha} & = & 
i\mathsf{U}_{\alpha I\, u}\not\!\!\mathfrak{D}q^{u}\epsilon^{I}
+N_{\alpha}{}^{I}\epsilon_{I}
=0\, ,
\end{eqnarray}

\noindent
admit at least one solution $\epsilon_{I}$, which is then called a \textit{Killing spinor}.  As
usual in this kind of analysis, we will not assume that the Bianchi identities
are satisfied by the field strengths of a configuration which should be
regarded as ``black boxes''. Imposing the Bianchi identities will be
equivalent to imposing that those black boxes are related to the potentials
(which are used explicitly in gauged theories) by
eq.~(\ref{eq:Fdef}); for the moment we will treat the
vectors $A^{\Lambda}{}_{\mu}$ and the vector field strengths
$F^{\Lambda}{}_{\mu\nu}$ as independent fields. We will impose the Bianchi
identities together with the equations of motion after we have found the
supersymmetric configurations and at the end we will have
supersymmetric solutions determined by the independent fields
$\left\{g_{\mu\nu}(x), A^{\Lambda}{}_{\mu}(x), Z^{i}(x), q^{u}(x)\right\}$.

We start by studying the integrability conditions of the above KSEs:
using the supersymmetry transformation rules of the bosonic fields
eqs.~(\ref{eq:susytranse}--\ref{eq:susytransZ}) and using the results of
refs.~\cite{Kallosh:1993wx,Bellorin:2005hy} we can derive the following KSIs
satisfied by any field configuration
%$\left\{g_{\mu\nu}(x), F^{\Lambda}{}_{\mu\nu}(x),A^{\Lambda}{}_{\mu}(x), Z^{i}(x),q^{u}(x)\right\}$ 
admitting Killing spinors:

\begin{eqnarray}
\label{eq:ksi1n2}
\mathcal{E}_{a}{}^{\mu}\gamma^{a}\epsilon^{I}
-4i\varepsilon^{IJ}\mathcal{L}^{\Lambda}
\mathcal{E}_{\Lambda}{}^{\mu} \epsilon_{J}
& = & 0\, ,
\\
& & \nonumber \\
\label{eq:ksi2n2}
\mathcal{E}^{i}\epsilon^{I} 
-2i\varepsilon^{IJ}f^{*\, i\Lambda}\not\!\mathcal{E}_{\Lambda} 
\epsilon_{J} & = & 0\, ,
\\
& & \nonumber \\
\label{eq:hyperKSI}
\mathcal{E}^{u}\ \mathsf{U}^{\alpha I}{}_{u} \epsilon_{I} 
& = &  0\,  ,  
\end{eqnarray}

The vector field Bianchi identities eq.~(\ref{eq:BL}) do not appear in these
relations because the procedure used to derive them assumes the existence of
the vector potentials, and therefore uses the vanishing of the Bianchi identities.

It is convenient to treat the Maxwell equations and Bianchi identities on an
equal footing as to preserve the symplectic covariance of the theory,
which means having a formally electric-magnetic duality-covariant version of
the above KSIs. This version can be found by performing duality rotations on the above
identities or from the integrability conditions of the KSEs (see {\em
  e.g.\/} \cite{Meessen:2006tu}).
% \begin{eqnarray}
% 4\gamma^{\nu}\mathfrak{D}_{[\mu|}\delta_{\epsilon}\psi_{I\, |\nu]}  
% & = & 0\, ,
% \\
% & & \nonumber \\
% -i\not\!\!\mathfrak{D} \delta_{\epsilon}\lambda^{Ii} 
% & = & 
% 0\, ,
% \\
% & & \nonumber \\
% \not\!\!\mathfrak{D} \delta_{\epsilon}\zeta_{\alpha}
% & = & 0\, ,  
% \end{eqnarray}
% \noindent
Both procedures give exactly the same symplectic-invariant
result, namely

\begin{eqnarray}
\label{eq:ksipsi}
\mathcal{E}_{a}{}^{\mu}\gamma^{a}\epsilon_{I} -4i
\langle\, \mathcal{E}^{\mu} \mid \, \mathcal{V}\, \rangle 
\varepsilon_{IJ}\epsilon^{J} & = & 0\, ,\\
& & \nonumber \\
\label{eq:ksilambda} 
\mathcal{E}^{i} \epsilon^{I} +2i
\langle\, \not\!\mathcal{E} \mid \, \mathcal{U}^{*\, i}\, \rangle 
\varepsilon^{IJ} \epsilon_{J}& = & 0\, ,
\\
& & \nonumber \\
\label{eq:ksihyper}
\mathcal{E}^{u}\ \mathsf{U}^{\alpha I}{}_{u} \epsilon_{I} 
& = &  0\, ,  
\end{eqnarray}

\noindent
where $\mathcal{E}^{\mu}$ is a symplectic vector containing the Maxwell
equations and Bianchi identities 

\begin{equation}
\mathcal{E}^{\mu}
\equiv
\left(
  \begin{array}{c}
\mathcal{B}^{\Lambda\,  \mu}\\ \mathcal{E}_{\Lambda}{}^{\mu} \\
\end{array}
\right)\, . 
\end{equation}

Acting on these identities with gamma matrices and conjugate spinors from the
left, we get identities involving the equations of motion and tensors which
are the bilinears of the Killing spinors. 
As mentioned in the introduction, there are two cases to be considered,
the sexer being the causal nature of the vector bilinear $V^{a}\equiv
i\bar{\epsilon}^{I}\gamma^{a}\epsilon_{I}$, namely the timelike and
the null case. In the
timelike case (the only one we are going to consider in this paper) we can use
$V^{a}/|V|$ as the component $e^{0}$ of an orthonormal frame, obtaining the
identities

\begin{eqnarray}
\mathcal{E}^{0m} 
& = & 
\mathcal{E}^{mn}=0\, ,
\\
& & \nonumber \\
\langle\, \mathcal{V}/X \mid \, \mathcal{E}^{0} \, \rangle  
& = & 
{\textstyle\frac{1}{4}}|X|^{-1} \mathcal{E}^{00}\, ,
\\
& & \nonumber \\
\langle\, \mathcal{V}/X \mid \, \mathcal{E}^{m} \, \rangle  
& = & 
0\, ,
\\
& & \nonumber \\
\langle\, \mathcal{U}^{*}_{i^{*}}\mid \, \mathcal{E}^{0} \, \rangle  
& = & 
{\textstyle\frac{1}{2}} e^{-i\alpha}\mathcal{E}_{i^{*}}\, ,
\\
& & \nonumber \\
\langle\, \mathcal{U}^{*}_{i^{*}}\mid \, \mathcal{E}^{m} \, \rangle  
& = & 
0\, ,
\\
& & \nonumber \\
\mathcal{E}^{u}  
& = & 
0\, ,
\end{eqnarray}

\noindent
where $X \equiv\tfrac{1}{2}\varepsilon^{IJ}\bar{\epsilon}_{I}\epsilon_{J}$ is
the scalar bilinear and $\alpha$ is its phase
\cite{Bellorin:2005zc}. These identities imply that \cite{Gauntlett:2002nw,Bellorin:2005hy}
\begin{enumerate}
\item All the supersymmetric configurations automatically satisfy all the
  equations of motion except for $\mathcal{E}^{0}$ and
  $\mathcal{E}^{00}$ and also the Bianchi identities.
\item We will only need to impose $\mathcal{E}^{0}=0$ on the supersymmetric
  configurations in order to have a solution of all the classical equations of
  motion and Bianchi identities.
\end{enumerate}

% Further, the r.h.s.~of eq.~(\ref{eq:ksi1n2}) is real, and this leads to
% two important identities:

% \begin{eqnarray}
% \langle\, \mathcal{I} \mid \, \mathcal{E}^{0} \, \rangle  & = & 0\, ,\\
% & & \nonumber \\
% \label{eq:ksi2-1}
% \mathcal{E}^{00} & = & \pm 4|\langle\, \mathcal{V} \mid \, \mathcal{E}^{0}\,
% \rangle|\, .
% \end{eqnarray}
%
%%%%%%%%%%%%%%%%%%%%%%%%%%%%%%%%%%%%%%%%%%%%%%%%%%%%%%%%%%%%%%%%%%%%%%
%%%%%%%%%%%%%%%%%%%%%%%%%%%%%%%%%%%%%%%%%%%%%%%%%%%%%%%%%%%%%%%%%%%%%%
%%%%%%%%%%%%%%%%%%%%%%%%%%%%%%%%%%%%%%%%%%%%%%%%%%%%%%%%%%%%%%%%%%%%%%
%%%%%%%%%%%%%%%%%%%%%%%%%%%%%%%%%%%%%%%%%%%%%%%%%%%%%%%%%%%%%%%%%%%%%%
\section{Killing equations for the bilinears}
\label{sec-bilinearkse}
%%%%
{}From the gravitino supersymmetry transformation rule
eq.~(\ref{eq:gravisusyrulehyper}), using the decompositions
eqs.~(\ref{eq:vectordecomposition})-(\ref{eq:2formdecomposition2}) we get the
independent equations

\begin{eqnarray}
\label{eq:DXVT}
\mathfrak{D}_{\mu}X 
& = & 
i V^{\nu}T^{+}{}_{\nu\mu}
+{\textstyle\frac{i}{\sqrt{2}}}S^{x}V^{x}{}_{\mu}\, ,
\\
& & \nonumber \\
\label{eq:VKilling}
\nabla_{(\mu}V_{\nu)} 
& = & 0\, ,
\\
& & \nonumber \\
\label{eq:dVhyper}
d\hat{V} 
& = & 
4iX\hat{T}^{*-} -\sqrt{2}S^{*x}\hat{\Phi}^{x}+\mathrm{c.c.}\, ,
\\  
& & \nonumber \\
\label{eq:VxKilling}
\mathfrak{D}_{(\mu} V^{x}{}_{\nu)} 
& = & 
T^{*-}{}_{(\mu|\rho}\Phi^{x}{}_{|\nu)}{}^{\rho} 
+{\textstyle\frac{i}{\sqrt{2}}}XS^{*x}g_{\mu\nu}
+\mathrm{c.c.}\, ,
\\
& & \nonumber \\
\label{eq:DVx}
\mathfrak{D}\hat{V}^{x} 
& = & 
-i\epsilon^{xyz}S^{*y}\hat{\Phi}^{z}+\mathrm{c.c.}\, ,
\end{eqnarray}

\noindent
where we denote differential forms with hats, and the $\mathrm{SU}(2)$-covariant
derivative is

\begin{equation}
\mathfrak{D}\hat{V}^{x}
=
d\hat{V}^{x}
+\epsilon^{xyz}\hat{\mathsf{A}}^{y} \wedge \hat{V}^{z} \, .
\end{equation}

Eq.~(\ref{eq:VKilling}) indicates that $V$ is, as usual in SUGRA, a timelike
Killing vector.  According to eq.~(\ref{eq:VxKilling}), the vectors $V^{x}$
are not, in general. However, for vanishing graviphoton field strength, they
are conformal Killing vectors. The equations for $d\hat{V}$ (\ref{eq:dVhyper})
and $\mathfrak{D}\hat{V}^{x}$ (\ref{eq:DVx}) will be used and analyzed later
on.

{}From the gauginos' supersymmetry transformation rules,
eqs.~(\ref{eq:gaugsusyrulehyper}), we get

\begin{eqnarray}
0 & =& V^{I}{}_{K}{}^{\mu}\mathfrak{D}_{\mu}Z^{i} +\varepsilon^{IJ}\Phi_{KJ}{}^{\mu\nu}
G^{i\, +}{}_{\mu\nu} +W^{i}\delta^{I}{}_{K}
+XW^{i\,I}{}_{K} \, ,\\
& & \nonumber \\
0 & =& iX^{*}\varepsilon^{KI}\mathfrak{D}^{\mu}Z^{i} 
 +i\Phi^{KI\, \mu\nu}\mathfrak{D}_{\nu}Z^{i}
-4i\varepsilon^{IJ}G^{i\, +\, \mu}{}_{\nu}V^{K}{}_{J}{}^{\nu} 
\nonumber \\
& & \nonumber \\
& & -i W^{i}\varepsilon^{IJ}V^{K}{}_{J}{}^{\mu}-iW^{i\,IJ}V^{K}{}_{J\, \mu} 
\, .
\end{eqnarray}

\noindent
The trace of the first equation gives

\begin{equation}
\label{eq:VdZhyper}
V^{\mu}\mathfrak{D}_{\mu}Z^{i}+2X W^{i}=0\, ,  
\end{equation}

\noindent
while the antisymmetric part of the second equation gives 

\begin{equation}
\label{eq:GVhyper}
2 X^{*}\mathfrak{D}_{\mu}Z^{i}+4 G^{i\, +}{}_{\mu\nu}V^{\nu} 
+W^{i} V_{\mu}-W^{i\ J}{}_{K} V^{K}{}_{J\, \mu} =0\, .  
\end{equation}

{}From eqs.~(\ref{eq:DXVT}) and (\ref{eq:GVhyper}) we get 

\begin{eqnarray}
\label{eq:VT+}
V^{\nu}T^{+}{}_{\nu\mu} 
& = &
-i\mathfrak{D}_{\mu}X 
-{\textstyle\frac{1}{\sqrt{2}}} S^{x}V^{x}{}_{\mu}\, ,\\ 
& & \nonumber \\
\label{eq:VGi+}
V^{\nu}G^{i\, +}{}_{\nu\mu} 
& = & 
{\textstyle\frac{1}{2}} X^{*}\mathfrak{D}_{\mu}Z^{i}  
+{\textstyle\frac{1}{4}} W^{i}V_{\mu}
-{\textstyle\frac{i}{4\sqrt{2}}} W^{ix}V^{x}{}_{\mu}\, .
\end{eqnarray}

\noindent
The consistency of these expressions requires

\begin{equation}
\label{eq:VdXhyper}
V^{\mu}\mathfrak{D}_{\mu}X \; =\; 0\, ,  
\end{equation}

\noindent
and eq.~(\ref{eq:VdZhyper}), respectively. Upon using the Special
Geometry completeness relation \cite{Andrianopoli:1996cm}

\begin{equation}
\label{eq:completeness}
F^{\Lambda\, +}=
i\mathcal{L}^{*\, \Lambda}T^{+} +2f^{\Lambda}{}_{i}G^{i\, +}\, ,  
\end{equation}

\noindent
we obtain from eqs.~(\ref{eq:VT+}) and (\ref{eq:VGi+}), first of all 

\begin{equation}
\label{eq:VFLup+}
V^{\nu}F^{\Lambda\, +}{}_{\nu\mu} 
=  
\mathcal{L}^{*\, \Lambda}\mathfrak{D}_{\mu}X
+X^{*}\mathfrak{D}_{\mu} \mathcal{L}^{\Lambda} 
+{\textstyle\frac{i}{8}}g \Im{\rm m}\mathcal{N}^{-1|\Lambda\Sigma} 
(\mathcal{P}_{\Sigma}V_{\mu} +\sqrt{2}\, \mathsf{P}_{\Sigma}{}^{x}
V^{x}{}_{\mu})\, .
\end{equation}

\noindent
Then, using $F_{\Lambda}{}^{+}= \mathcal{N}^{*}_{\Lambda\Sigma}F^{\Sigma\,
  +}$, we get for the symplectic vector of field strengths
$\mathcal{F}^{T}\equiv \left(\ F^{\Lambda}, F_{\Lambda}\right)$
% \left(
%   \begin{array}{c}
% F^{\Lambda} \\ F_{\Lambda} \\
% \end{array}
% \right)$

\begin{equation}
\label{eq:VF+}
V^{\nu}\mathcal{F}^{+}{}_{\nu\mu}   
=
\mathcal{V}^{*}\mathfrak{D}_{\mu}X
+X^{*}\mathfrak{D}_{\mu} \mathcal{V} 
\ -\ {\textstyle\frac{i}{8}}g\, \Omega^{-1} \left(\mathcal{M} +i\Omega\right) 
\ \left[\mathcal{P}V_{\mu} +\sqrt{2}\mathsf{P}^{x}
V^{x}{}_{\mu}\right]\, ,
\end{equation}

\noindent
where $\mathcal{M}$ and $\Omega$ are the symplectic matrices

\begin{equation}
\mathcal{M}
\equiv
\left(
\begin{array}{cc}
I+RI^{-1}R  & -RI^{-1} \\
& \\
-I^{-1}R & I^{-1} \\   
\end{array}
\right)\, ,
\hspace{.5cm}
\Omega
\equiv
\left(
\begin{array}{cc}
0  & \mathbb{I} \\
-\mathbb{I} & 0 \\   
\end{array}
\right)\, ,
\end{equation}

\noindent
and we have defined

\begin{equation}
I_{\Lambda\Sigma}
\equiv 
\Im{\rm m}\, (\mathcal{N}_{\Lambda\Sigma})\, ,
\hspace{.5cm}
R_{\Lambda\Sigma}
\equiv 
\Re{\rm e}\, (\mathcal{N}_{\Lambda\Sigma})\, ,
\hspace{.5cm}
I^{\Lambda\Sigma}I_{\Sigma\Omega} = \delta^{\Sigma}{}_{\Omega}\, ;  
\end{equation}

\noindent
furthermore, we have introduced the following symplectic vectors for the momentum maps

\begin{equation}
\mathcal{P} \ =\
\left(
  \begin{array}{c}
 0 \\ \mathcal{P}_{\Lambda} \\   
  \end{array}
\right)\, ,  
\hspace{1cm}
\mathsf{P}^{x}\ =\
\left(
  \begin{array}{c}
 0 \\ \mathsf{P}_{\Lambda}{}^{x} \\   
  \end{array}
\right)\, .  
\end{equation}

Had we used the embedding-tensor formalism \cite{Cordaro:1998tx}, none of the components of these
symplectic vectors would have vanished and we would have obtained manifestly
symplectic-invariant expressions; using only the fundamental (electric) 1-forms
as gauge fields, however, kills off half of the components, as we have seen above.

After some straightforward manipulations we obtain the general form of the
electric and magnetic field strengths

\begin{eqnarray}
\label{eq:TimeFfinal}
\mathcal{F} 
& = & 
-\tfrac{1}{2}\ 
\mathfrak{D}[ \mathcal{R}\ \hat{V}]
\ +\ 
\frac{g}{8\sqrt{2} |X|^{2}}\ \mathsf{P}^{x}\ \hat{V}\wedge\hat{V}^{x} 
                   \nonumber \\
& & \nonumber \\
& & 
-\tfrac{1}{2}
\star\left\{
            \hat{V}\wedge\left[\
               \mathfrak{D}\mathcal{I} 
               \ -\ \sqrt{2}g\ \left(\
                      \mathcal{R}\ \langle\mathcal{R}|\mathsf{P}^{x}\rangle
                      \ -\ \frac{1}{8|X|^{2}}\ \Omega^{-1}\mathcal{M}\ \mathsf{P}^{x}
                    \right)\ \hat{V}^{x}\
            \right]\right\}\, ,
\end{eqnarray}

\noindent
where following ref.~\cite{Meessen:2006tu} we have defined 

\begin{equation}
\label{eq:SeedVector}
\mathcal{V}/X \;\equiv\; \mathcal{R}\; +\; i\mathcal{I}\, .  
\end{equation}

Let us now consider the hyperini's KSE: it is convenient to rewrite it as

\begin{equation}
\label{eq:KSE2N3-2}
\not\!\!\mathfrak{D}q^{u}\epsilon^{I} 
-i\mathsf{K}^{x\, u}{}_{v}\sigma^{x}{}_{J}{}^{I} 
\not\!\!\mathfrak{D}q^{v}\epsilon^{J}
-ig \varepsilon^{IJ}\mathcal{L}^{*\, \Lambda}
\mathsf{k}_{\Lambda}{}^{u}\epsilon_{J}
+{\textstyle\frac{1}{2}}g\mathcal{L}^{*\, \Lambda}
\mathfrak{D}^{u}\mathsf{P}_{\Lambda}{}^{x}\sigma^{x\, IJ}\epsilon_{J}
=  
0\, .
\end{equation}

We only get one independent equation for the bilinears:

\begin{equation}
\label{eq:VIKDq}
V^{I}{}_{K}{}^{\mu}\mathfrak{D}_{\mu}q^{u}
-i\mathsf{K}^{x\, u}{}_{v}\sigma^{x}{}_{J}{}^{I}
V^{J}{}_{K}{}^{\mu}\mathfrak{D}_{\mu}q^{v}
+g X\delta^{I}{}_{K}\mathcal{L}^{*\, \Lambda}
\mathsf{k}_{\Lambda}{}^{u}  
+{\textstyle\frac{i}{2}}gX\mathcal{L}^{*\, \Lambda}
\mathfrak{D}^{u}\mathsf{P}_{\Lambda}{}^{x}\sigma^{x\, I}{}_{K}
=
0\, .
\end{equation}

\noindent
The trace of this equation is

\begin{equation}
V^{\mu}\mathfrak{D}_{\mu}q^{u}
-i\sqrt{2}\mathsf{K}^{x\, u}{}_{v}V^{x\, \mu}\mathfrak{D}_{\mu}q^{v}
+2g X\mathcal{L}^{*\, \Lambda}\mathsf{k}_{\Lambda}{}^{u}  
=
0\, ,
\end{equation}

\noindent
and its real and imaginary parts are

\begin{eqnarray}
\label{eq:VDq}
V^{\mu}\mathfrak{D}_{\mu}q^{u}
+2g |X|^{2}\mathcal{R}^{\Lambda}\mathsf{k}_{\Lambda}{}^{u}  
& = & 
0\, ,\\
& & \nonumber \\
\label{eq:DP}
\mathsf{K}^{x\, u}{}_{v}V^{x\, \mu}\mathfrak{D}_{\mu}q^{v}
+\sqrt{2}g |X|^{2}\mathcal{I}^{\Lambda}\mathsf{k}_{\Lambda}{}^{u}  
& = & 
0\, .
\end{eqnarray}

The rest of the equations that can be obtained from 
eq.~(\ref{eq:VIKDq}) can also be obtained from these two. In particular, we
can get from eq.~(\ref{eq:DP}) 

\begin{equation}
\label{eq:VxDq}
V^{x\, \mu}\mathfrak{D}_{\mu}q^{u}
+\varepsilon_{xyz}\mathsf{K}^{y\, u}{}_{v}V^{z\, \mu}\mathfrak{D}_{\mu}q^{v}
+{\textstyle\frac{1}{\sqrt{2}}}
g |X|^{2}\mathcal{I}^{\Lambda}\mathfrak{D}^{u}\mathsf{P}_{\Lambda}{}^{x}  
=0\, .  
\end{equation}

In order to make further progress we must introduce coordinates and
obtain
information about the metric.
%%%%%%%%%%%%%%%%%%%%%%%%%%%%%%%%%%%%%%%%%%%%%%%%%%%%%%%%%%%%%%%%%%%%%%
%%%%%%%%%%%%%%%%%%%%%%%%%%%%%%%%%%%%%%%%%%%%%%%%%%%%%%%%%%%%%%%%%%%%%%
%%%%%%%%%%%%%%%%%%%%%%%%%%%%%%%%%%%%%%%%%%%%%%%%%%%%%%%%%%%%%%%%%%%%%%
%%%%%%%%%%%%%%%%%%%%%%%%%%%%%%%%%%%%%%%%%%%%%%%%%%%%%%%%%%%%%%%%%%%%%%
\section{The metric} 
\label{sec-timelikemetric}
We define a time coordinate $t$ associated to the timelike Killing vector $V$
by
\begin{equation}
\label{eq:t-timelike}
V^{\mu}\partial_{\mu}\; \equiv\; \sqrt{2} \partial_{t}\, .
\end{equation}

\noindent
Then, by choosing the gauge fixing condition

\begin{equation}
\label{eq:gaugechoice}
V^{\mu}A^{\Lambda}{}_{\mu} =\sqrt{2}A^{\Lambda}{}_{t}= 
-2|X|^{2}\mathcal{R}^{\Lambda}\, ,  
\end{equation}

\noindent 
we can solve eqs.~(\ref{eq:VdZhyper},\ref{eq:VdXhyper}) and (\ref{eq:VDq}) by
taking all the scalar fields and the function $X$ to be
time-independent,\footnote{The consistency of this gauge choice in all the
  equations derived from the KSEs requires the use of several identities that
  can be derived from the generic expression of the momentum map
  $\mathcal{P}_{\Lambda}$ eq.~(\ref{eq:1}), which is equivalent to
\begin{equation}
 \mathcal{P}_{\Lambda}= 2|X|^{2}f_{\Lambda\Sigma}{}^{\Omega}
\left(\mathcal{R}^{\Sigma}\mathcal{R}_{\Omega}
+\mathcal{I}^{\Sigma}\mathcal{I}_{\Omega}\right)\, ,   
\end{equation}
the property eq.~(\ref{eq:VSLV2}) and $\mathfrak{D}\mathcal{M}_{\Lambda} =
\mathcal{N}^{*}_{\Lambda\Sigma}\mathfrak{D}\mathcal{L}^{\Sigma}$. These
properties  are
\begin{eqnarray}
f_{\Lambda\Sigma}{}^{\Omega}\mathcal{R}^{\Sigma}\mathcal{R}_{\Omega}
& = & 
f_{\Lambda\Sigma}{}^{\Omega}\mathcal{I}^{\Sigma}\mathcal{I}_{\Omega}
=
I_{\Lambda\Gamma}f_{\Sigma\Omega}{}^{\Gamma}\mathcal{R}^{\Sigma}\mathcal{I}^{\Omega}\, ,
\\
& & \nonumber \\
f_{\Lambda\Sigma}{}^{\Omega}\mathcal{R}^{\Sigma}\mathcal{I}_{\Omega}
& = & 
-f_{\Lambda\Sigma}{}^{\Omega}\mathcal{I}^{\Sigma}\mathcal{R}_{\Omega}
=
-R_{\Lambda\Gamma}T^{\Gamma\Delta}f_{\Delta\Sigma}{}^{\Omega}
\mathcal{R}^{\Sigma}\mathcal{R}_{\Omega}\, .
\end{eqnarray}
}
{\em i.e.\/}

\begin{equation}
\label{eq:VdZVdXVdq0}
\partial_{t}Z^{i}=\partial_{t}X=\partial_{t}q^{u}=0\, .  
\end{equation}

The definition eq.~(\ref{eq:t-timelike}) and the Fierz identity
$V^{2}=4|X|^{2}$ imply that $\hat{V}$ must take the form

\begin{equation}
\label{eq:hatV} 
\hat{V}=2\sqrt{2}|X|^{2}(dt+\hat{\omega})\, , 
\end{equation}

\noindent
where $\hat{\omega}$ is a spatial 1-form, which by definition, must satisfy 

\begin{equation}
d\hat{\omega}=\tfrac{1}{2\sqrt{2}} d(|X|^{-2}\hat{V})\, .
\end{equation}

\noindent
eqs.~(\ref{eq:dVhyper},\ref{eq:DXVT}) and some straightforward manipulations
imply that $\hat{\omega}$ must satisfy

\begin{equation}
\label{eq:dog}
d\hat{\omega}=
-\tfrac{i}{2\sqrt 2} \star 
\left[\left(X\mathfrak{D}X^*-X^*\mathfrak{D}X
\ +\ ig\sqrt{2} |X|^2\mathcal{R}^{\Lambda}
\mathsf{P}_{\Lambda}{}^{x}\hat{V}^{x}\right)
\wedge\frac{\hat{V}}{|X|^{4}}\right]\, .
\end{equation}

% With this choice of time coordinate, the scalars $Z^{i},q^{u},X$ are
% time-independent according to eqs.~(\ref{eq:VdZVdXVdq0}).

Since the $\hat{V}^{x}$s are not exact, we cannot simply define coordinates by
putting $\hat{V}^{x}\equiv dx^{x}$. We can, however, still use them to
construct the metric: using

\begin{equation}
g_{\mu\nu}=2V^{-2}[V_{\mu}V_{\nu}-V^{J}{}_{I\, \mu} V^{I}{}_{J\, \nu}]\, ,
\end{equation}

\noindent
and the decomposition eq.~(\ref{eq:vectordecomposition}),
we find that the metric can be written in the form

\begin{equation}
ds^{2}= \frac{1}{4|X|^{2}}\hat{V}\otimes \hat{V} 
-\frac{1}{2|X|^{2}} \delta_{xy}\hat{V}^{x}\otimes \hat{V}^{y}\, .
\end{equation}

\noindent
The $\hat{V}^{x}$ are mutually orthogonal and also orthogonal to $\hat{V}$,
which means that they can be used as a Dreibein for a 3-dimensional Euclidean
metric

\begin{equation}
\delta_{xy}\hat{V}^{x}\otimes \hat{V}^{y} \; \equiv\; 
\gamma_{\underline{m}\underline{n}}dx^{m}dx^{n}\, ,  
\end{equation}

\noindent
where we introduced the remaining 3 spatial coordinates $x^{m}$ ($m=1,2,3$).
The 4-dimensional metric takes the coordinate-form

\begin{equation}
\label{eq:timelikemetric}
ds^{2} \; =\; 2|X|^{2} (dt+\hat{\omega})^{2} 
-\frac{1}{2|X|^{2}}\gamma_{\underline{m}\underline{n}}dx^{m}dx^{n}\, .  
\end{equation}

In what follows we will use the Vierbein basis

\begin{equation}
e^{0}  =  \frac{1}{2|X|} \hat{V}\, ,
\hspace{1cm}
e^{x}  =  \frac{1}{\sqrt{2}|X|}\hat{V}^{x}\, ,
\end{equation}

\noindent
that is

\begin{equation}
\label{eq:Vierbeins}
(e^{a}{}_{\mu}) = 
\left(
  \begin{array}{cc}
\sqrt{2}|X| & \sqrt{2}|X| \omega_{\underline{m}} \\
& \\
0 & \frac{1}{\sqrt{2}|X|} V^{x}{}_{\underline{m}} \\
  \end{array}
\right)\, ,
\hspace{1cm}
(e^{\mu}{}_{a}) = 
\left(
  \begin{array}{cc}
\frac{1}{\sqrt{2}|X|} & -\sqrt{2}|X| \omega_{x} \\
& \\
0 & \sqrt{2}|X| V_{x}{}^{\underline{m}} \\
  \end{array}
\right)\, .
\end{equation}

\noindent
where $V_{x}{}^{\underline{m}}$ is the inverse Dreibein
$V_{x}{}^{\underline{m}} V^{y}{}_{\underline{m}}=\delta^{y}{}_{x}$ and
$\omega_{x} = V_{x}{}^{\underline{m}}\omega_{\underline{m}}$.  Observe that we
can raise and lower flat 3-dimensional indices with $\delta_{xy}$ and $\delta^{xy}$,
whence their position is rather irrelevant.  We shall also adopt the
convention that, from now on, all objects with flat or curved 3-dimensional
indices refer to the above Dreibein and the corresponding metric. 
\par
Using these conventions, we see that eq.~(\ref{eq:dog}) takes the 3-dimensional form
\begin{equation}
\label{eq:do1}
(d\hat{\omega})_{xy}
=
-\frac{1}{2|X|^{2}} 
\varepsilon_{xyz}\left\{i \left(\frac{\tilde{\mathfrak{D}}_{z}X}{X}
-\frac{\tilde{\mathfrak{D}}_{z}X^{*}}{X^{*}} \right)
-\sqrt{2}g\, 
\langle\, \mathcal{R} \mid \mathsf{P}^{z}\, \rangle \right\}\, ,
\end{equation}

\noindent
or using the symplectic vectors defined in eq.~(\ref{eq:SeedVector})

\begin{equation}
\label{eq:do}
(d\hat{\omega})_{xy}\ =\
2\varepsilon_{xyz}\left\{
  \langle\mathcal{I}\mid \tilde{\mathfrak{D}}_{z}\mathcal{I}\rangle
    \ +\ \frac{g}{2\sqrt{2}|X|^{2}}
\langle\, \mathcal{R} \mid \mathsf{P}^{z}\, \rangle \right\}\, ,
\end{equation}

\noindent
where $\tilde{\mathfrak{D}}$ is the covariant derivative w.r.t.~the effective
3-dimensional gauge connection

\begin{equation}
\label{eq:effectivegaugeconnection}
\tilde{A}^{\Lambda}{}_{\underline{m}} \ \equiv\ 
A^{\Lambda}{}_{\underline{m}}  -\omega_{\underline{m}}A^{\Lambda}{}_{t}
\ =\
A^{\Lambda}{}_{\underline{m}}  
+\sqrt{2}|X|^{2}\mathcal{R}^{\Lambda}\omega_{\underline{m}}\, .
\end{equation}

Let us now consider eq.~(\ref{eq:DVx}): the mixed indices part takes
on the form, using the gauge fixing eq.~(\ref{eq:gaugechoice}), 

\begin{equation}
\partial_{t}V^{x}{}_{\underline{m}} = 0\, ,
\end{equation}

\noindent
while the purely spatial part takes the form

\begin{equation}
d\hat{V}^{x} +\epsilon^{xyz} \tilde{\hat{A}}^{y}\wedge \hat{V}^{z}
+\hat{T}^{x}=0\, ,  
\end{equation}

\noindent
where

\begin{eqnarray}
\label{eq:effectivesu2connection}
\tilde{\hat{A}}^{x}{}_{\underline{m}} 
& \equiv & 
\mathsf{A}^{x}{}_{\underline{m}} 
-\sqrt{2}g |X|^{2} \, 
\langle\, \tilde{A}_{\underline{m}} \mid \mathsf{P}^{x}\, \rangle \omega_{\underline{m}}
=                                                                                  
\mathsf{A}^{x}{}_{\underline{m}} 
-g\, \langle\, \tilde{A}_{\underline{m}} \mid \mathsf{P}^{x}\, \rangle \, , 
\\
& & \nonumber \\
\hat{T}^{x} & = & 
-{\textstyle\frac{1}{2\sqrt{2}}}g\epsilon^{xyz} 
\langle\, \mathcal{I} \mid \mathsf{P}^{y}\, \rangle\, 
\epsilon^{zvw}\hat{V}^{v}\wedge\hat{V}^{w}\, .  
\end{eqnarray}

The above equation can be interpreted as Cartan's first structure
equations for the Dreibein $\hat{V}^{x}$, the $\mathrm{SU}(2)$-connection
1-form $\tilde{\hat{A}}^{x}$ and the torsion 2-form $\hat{T}^{x}$. It can be
solved for the spin connection as a function of the Dreibein and
torsion, {\em i.e.\/}

\begin{equation}
\label{eq:embedding}
 \varpi_{xyz}(V) =-\varepsilon_{yzw}\tilde{A}^{w}{}_{x}  - K_{xyz}(T)\, ,  
\end{equation}

\noindent
where $\varpi_{xyz}(V)$ is the standard 3-dimensional Levi-Civit\`a connection
1-form (which is completely determined by the Dreibein), and $K_{xyz}(T)$ is the
contorsion 1-form, to wit

\begin{equation}
\label{eq:contorsion}
K_{xyz} = {\textstyle\frac{1}{2}} \{ T_{xzy}+T_{yzx}-T_{xyz}\}
=
-\sqrt{2}g\ \langle\, \mathcal{I} \mid \mathsf{P}^{[y}\, \rangle\,  \delta^{z]x}\, .  
\end{equation}

This condition relates the spin connection of the 3-dimensional space with the
pullback of the $\mathrm{SU}(2)$-connection, the gauge connection and the
complex scalars. In the ungauged case, considered in
ref.~\cite{Huebscher:2006mr}, this complicated relation reduces to a
straightforward relation between the first two.

Let us summarize our results: we have shown that

\begin{enumerate}
\item The metric of a bosonic field configuration of $N=2$, $d=4$ supergravity
  $A^{\Lambda}{}_{\mu},Z^{i},q^{u}$ can be written in the conformastationary
  form eq.~(\ref{eq:timelikemetric}) where the spatial 1-form $\hat{\omega}$
  satisfies eq.~(\ref{eq:do}) and the spin connection of spatial 3-dimensional
  metric $\gamma_{\underline{m}\underline{n}}$ is related to the pullback of
  the quaternionic-K\"ahler $\mathrm{SU}(2)$ connection $\mathsf{A}^{I}{}_{J\, \mu}$
  and the gauge connection by
  eqs.~(\ref{eq:effectivesu2connection},\ref{eq:effectivegaugeconnection})
  and (\ref{eq:embedding}).
\item The vector field strengths must take the form that can be derived from
  eq.~(\ref{eq:VFLup+}).
\item The covariant derivatives of the hyperscalars must satisfy
  eqs.~(\ref{eq:VDq}) and (\ref{eq:DP}). In the gauge
  eq.~(\ref{eq:gaugechoice}), eq.~(\ref{eq:VDq}) just states that the
  hyperscalars are time-independent. 
\item The complex scalars $Z^{i}$ must satisfy eq.~(\ref{eq:VdZhyper})
  and in the
  gauge eq.~(\ref{eq:gaugechoice}) they are also time-independent. Observe
  that there are no further equations for them.
\end{enumerate}

In the next section we are going to show that the necessary conditions that we
have just found are also sufficient to have unbroken supersymmetry.
%%%%%%%%%%%%%%%%%%%%%%%%%%%%%%%%%%%%%%%%%%%%%%%%%%%%%%%%%%%%%%%%%%%%%%
%%%%%%%%%%%%%%%%%%%%%%%%%%%%%%%%%%%%%%%%%%%%%%%%%%%%%%%%%%%%%%%%%%%%%%
%%%%%%%%%%%%%%%%%%%%%%%%%%%%%%%%%%%%%%%%%%%%%%%%%%%%%%%%%%%%%%%%%%%%%%
%%%%%%%%%%%%%%%%%%%%%%%%%%%%%%%%%%%%%%%%%%%%%%%%%%%%%%%%%%%%%%%%%%%%%%
%%%%%%%%%%%%%%%%%%%%%%%%%%%%%%%%%%%%%%%%%%%%%%%%%%%%%%%%%%%%%%%%%%%%%%
\section{Killing spinor equations: necessary is also sufficient}
\label{sec:suffi}
%%%%
Let us consider first the gaugini KSE eq.~(\ref{eq:KSE2N2}):
by straightforwardly expanding and manipulating the ingredients
one can put it in the following form

\begin{equation}
\label{eq:KSENisS1}
\delta_{\epsilon}\lambda^{Ii} \; =\; 
i\sqrt{2}|X|\gamma^{x}\tilde{\mathfrak{D}}_{x}Z^{i}\ \left(\Pi^{0}\epsilon\right)^{I}
      \ -\ ie^{i\alpha}W^{i}\gamma^{0}\ \left(\Pi^{0}\epsilon\right)^{I}
      \ -\ iW^{ix}\gamma^{0x}\varepsilon^{IL}\ \Pi^{x}{}_{L}{}^{K}\ \epsilon_{K}\ ,
\end{equation}

\noindent
where we have  defined, as was indicated before, $X=e^{i\alpha}\ |X|$
and
%
% \begin{equation}
% X = e^{i\alpha}\ |X|\, ,
% \end{equation}
% \noindent
% and 
%
\begin{eqnarray}
\label{eq:KSENisS2}
\left(\Pi^{0}\epsilon\right)^{I}\; 
& \equiv\; & 
\epsilon^{I}\; +\; ie^{-i\alpha}\gamma^{0}\ \varepsilon^{IJ}\epsilon_{J} \; ,
\\
& & \\
\label{eq:KSENisS3}
\Pi^{x\, I}{}_{J}
& \; =\; &
\tfrac{1}{2}
\left[
\delta^{I}{}_{J} \ +\ \gamma^{0(x)}(\sigma^{(x)})^{I}{}_{J}
\right] \hspace{1cm}\mbox{(no sum)}\, .
\end{eqnarray}

The gaugini KSE, then, will be solved if we impose the projections

\begin{equation}
\left(\Pi^{0}\epsilon\right)^{I}=0\, ,
\hspace{1cm}
\Pi^{x\, I}{}_{J}
\epsilon_{I}=0\, ,  
\end{equation}

\noindent
for all $x$ for which $W^{ix}\neq 0$. The crucial properties of the $\Pi^{x}$
are

\begin{equation}
\label{eq:KSENisS4}
(\Pi^{x})^{2} \; =\; \Pi^{x} \;\; ,\;\;
  \mathrm{Tr}\left(\Pi^{x}\right) \; =\; 4 \hspace{1cm}\mbox{and}\hspace{1cm}
  \left[\ \Pi^{x}\ ,\ \Pi^{y}\ \right] \; =\; 0 \; ,
\end{equation}

\noindent
which guarantees that all 3 constraints $\Pi^{x\, I}{}_{J}\epsilon_{I}=0$ can,
if necessary, be consistently imposed at the same time. Furthermore, the
properties of the Pauli matrices (see Appendix~\ref{sec-sigmas})\footnote{
  This is easily seen to be
  true by making use of the identity $\varepsilon^{IL}\ \gamma^{0}\
  \Pi^{x}_{L}{}^{K} = \Pi^{xI}{}_{L}\ \gamma^{0}\ \varepsilon^{LK}$
  which expresses the fact that for the $\Pi^{x}$ complex conjugation is not
  the same as raising and lowering $SU(2)$ indices.
} 
ensure that these constraints are consistent with the fourth
constraint, namely $\left(\Pi^{0}\epsilon\right)^{I}=0$.

Having identified the pertinent projection operators, the remaining checks
of the KSEs are straightforward and we will be brief: the analysis of the 
hyperini variation (\ref{eq:KSE3N2}) implies that eq.~(\ref{eq:DP}) must
be satisfied and the $0$-direction of the gravitino variation (\ref{eq:KSE1N2})
implies that the Killing spinors are time-independent.
The analysis in the spacelike directions of the gravitino variation is best
expressed in terms of the K\"ahler-weight zero spinor $\eta_{I}$ defined by
$\epsilon_{I}=X^{1/2}\eta_{I}$. The parts of said variation that do not
cancel straightforwardly are

\begin{equation}
  \label{eq:KSENisS6}
  0 \; =\; \partial_{x}\eta_{I} \; +\; 
     \left[\
         \tfrac{1}{2}
\varpi_{xzz'}\ \varepsilon^{yzz'}
         \ +\ \tilde{\hat{A}}^{y}{}_{x}
         \ -\ \tfrac{1}{\sqrt{2}}g\varepsilon^{xzy}\ \mathcal{I}^{\Lambda}\mathsf{P}_{\Lambda}{}^{z}
     \ \right]\ \tfrac{i}{2}\ (\sigma^{y})^{J}{}_{I}\eta_{J} \; .
\end{equation}

\noindent
The identification of the spin connection in eqs.~(\ref{eq:embedding}) and
(\ref{eq:contorsion}), however, implies that the second term on the right hand
side vanishes, whence preserved supersymmetry implies that $\eta_{I}$ is
constant.

Summarizing and reformulating the results in this section we see that the Killing spinor
takes on the form $\epsilon_{I}=X^{1/2}\eta_{I}$, where $\eta_{I}$ is a constant spinor
satisfying

\begin{equation}
  \label{eq:KSENisS7}
  \Pi^{x\, J}{}_{I}\ \eta_{J} \; =\; 0 \hspace{1cm}\mbox{and}\hspace{1cm}
  0 \; =\; \eta^{I} \ +\ i\gamma^{0}\ \varepsilon^{IJ}\eta_{J} \; ,
\end{equation}

\noindent
the last restriction being a rescaled version of the constraint (\ref{eq:KSENisS2}).
\par
Since we must generically impose 4 compatible projection operators,
each of which is able to project out half of the components of the
Killing spinor, and we have 8 supercharges at our disposal we naively
should conclude that we must end up with a solution that has no
supersymmetry whatsoever. That this is not the case is due to the
structure of the $\Pi^{x}$'s and the chirality of the Killing spinor: it is easy to see that if we impose any
pair of them, say $\left(\Pi^{1}\epsilon\right)_{I} =0$ and $\left(\Pi^{2}\epsilon\right)_{I}$, then the spinor
automatically satisfies the third one, {\em i.e.\/} $\left(\Pi^{3}\epsilon\right)_{I}
=0$. This then means that the configurations that we
have obtained are $\textstyle{1\over 8}$-BPS.
%%%%%%%%%%%%%%%%%%%%%%%%%%%%%%%%%%%%%%%%%%%%%%%%%%%%%%%%%%%%%%%%%%%%%%
%%%%%%%%%%%%%%%%%%%%%%%%%%%%%%%%%%%%%%%%%%%%%%%%%%%%%%%%%%%%%%%%%%%%%%
%%%%%%%%%%%%%%%%%%%%%%%%%%%%%%%%%%%%%%%%%%%%%%%%%%%%%%%%%%%%%%%%%%%%%%
%%%%%%%%%%%%%%%%%%%%%%%%%%%%%%%%%%%%%%%%%%%%%%%%%%%%%%%%%%%%%%%%%%%%%%
%%%%%%%%%%%%%%%%%%%%%%%%%%%%%%%%%%%%%%%%%%%%%%%%%%%%%%%%%%%%%%%%%%%%%%
\section{Timelike supersymmetric solutions}
\label{sec:SusySol}
%%%%
The KSIs imply that the supersymmetric configurations that satisfy the zeroth
components of the Maxwell equations and (the Hodge dual of) the Bianchi
identities solve all the equations of motion of the theory.

The zeroth component of (the Hodge dual of) the Bianchi identities is just
the Bianchi identity of the effective 3-dimensional field strength
$\tilde{F}^{\Lambda}{}_{xy}$ which has the following 3-dimensional expression:

\begin{equation}
\label{eq:081198}
\tilde{F}^{\Lambda}{}_{xy} 
= 
-\tfrac{1}{\sqrt{2}}\varepsilon_{xyz} 
\{\tilde{\mathfrak{D}}_{z}\mathcal{I}^{\Lambda} +g\mathcal{B}^{\Lambda}{}_{z}\}\, ,
\end{equation}

\noindent
where

\begin{equation}
\mathcal{B}^{\Lambda}{}_{z}\equiv 
\sqrt{2}\left[\mathcal{R}^{\Lambda}\mathcal{R}^{\Sigma} 
+\frac{1}{8|X|^{2}} I^{\Lambda\Sigma}\right]\mathsf{P}_{\Sigma}{}^{z}\, .
\end{equation}

The above equation is a generalization of the well-known Bogomol'nyi equation
of Yang-Mills theories to an (almost arbitrary) 3-dimensional background metric
$\gamma_{\underline{m}\underline{n}}$ and with an extra term. If we find
$\tilde{A}^{\Lambda}{}_{\underline{m}},\mathcal{I}^{\Lambda},B^{\Lambda}{}_{x}$
solving that equation, then we have found a
$\tilde{A}^{\Lambda}{}_{\underline{m}}$ that gives rise to the field strength
$\tilde{F}^{\Lambda}{}_{\underline{m}\underline{n}}$ with the form prescribed
by supersymmetry and the 3-dimensional Bianchi identity and, therefore, the
zeroth component of the 4-dimensional one, are automatically satisfied.

The integrability equation of the Bogomol'nyi equation is a generalization of
the gauge-covariant Laplace equation for the $\mathcal{I}^{\Lambda}$:

\begin{equation}
\tilde{\mathfrak{D}}^{2}\mathcal{I}^{\Lambda}
+g\tilde{\mathfrak{D}}_{x}\mathcal{B}^{\Lambda}{}_{x}=0\, .  
\end{equation}

\noindent
Observe that in the above equation the covariant derivatives not only contain
the gauge part acting on the $\Lambda$-indices, but also the spin connection
for the 3-dimensional base space, which is constrained by
eq.~(\ref{eq:embedding})\footnote{ Observe that the components of the Abelian
  Bianchi identity w.r.t.~a curved frame reads $\nabla_{[a}F_{bc]}=0$, the
  extension to a non-Abelian one being obvious.}. In the ungauged, Abelian
cases, the $\mathcal{I}^{\Lambda}$ are just harmonic functions on
$\mathbb{R}^{3}$.

Let us now consider the zeroth component of the Maxwell equations, which can
be written as a sort of Bianchi identity for the dual field strengths
$F_{\Lambda}$: a lengthy calculation shows that the equation of motion leads to

\begin{equation}
\label{eq:TimMEOMexpl}
-\tfrac{1}{\sqrt{2}}\varepsilon_{xyz}  \tilde{\mathfrak{D}}_{x}\tilde{F}_{\Lambda\, yz} 
 =  
 \tfrac{1}{2\sqrt{2}}g \varepsilon_{xyz}(d\hat{\omega})_{xy}\mathsf{P}_{\Lambda}^{z}
+
\tfrac{1}{2}g^{2}f_{\Lambda (\Omega}{}^{\Gamma}f_{\Delta )\Gamma}{}^{\Sigma}\  
 \mathcal{I}^{\Omega}\mathcal{I}^{\Delta}\mathcal{I}_{\Sigma}
\ +\ \frac{g^{2}}{4|X|^{2}}\mathcal{R}^{\Sigma}
\mathsf{k}_{\Lambda\, u}\mathsf{k}_{\Sigma}{}^{u}\; ,
\end{equation}

\noindent
where we have defined\footnote{$\tilde{F}_{\Lambda\, xy}$ is strictly given by
  this definition because there are no dual 1-forms $A_{\Lambda}$ in this
  formulation.}

\begin{eqnarray}
  \label{eq:DefBeth}
\tilde{F}_{\Lambda\, xy} 
&   \;\equiv\; &
-\tfrac{1}{\sqrt{2}}\varepsilon_{xyz}
\left\{
 \tilde{\mathfrak{D}}_{x}\mathcal{I}_{\Lambda}
\, +\  g\mathcal{B}_{\Lambda\, x}\right\}\, ,
\\
& & \nonumber \\
\mathcal{B}_{\Lambda\, x}
& \equiv & 
\sqrt{2}
\left[ 
\mathcal{R}_{\Lambda}\mathcal{R}^{\Sigma} 
\ +\ \frac{1}{8|X|^{2}}\  R_{\Lambda\Gamma}I^{\Gamma\Sigma}\right]
 \mathsf{P}_{\Sigma}^{x} \; .
\end{eqnarray}

\noindent
If we use eq.~(\ref{eq:do}), which defines the 1-form
$\hat{\omega}$ to express the equation, as much as possible, in terms of
$\mathcal{R}$ and $\mathcal{I}$, we get

\begin{equation}
\label{eq:TimMEOMexpl2}
\begin{array}{rcl}
-\tfrac{1}{\sqrt{2}}\varepsilon_{xyz}  \tilde{\mathfrak{D}}_{x}\tilde{F}_{\Lambda\, yz} 
& = & 
 \tfrac{1}{\sqrt{2}}g
\langle\mathcal{I}\mid \tilde{\mathfrak{D}}_{x}\mathcal{I}\rangle\ \mathsf{P}_{\Lambda}^{x}
+
\tfrac{1}{2}g^{2}f_{\Lambda (\Omega}{}^{\Gamma}f_{\Delta )\Gamma}{}^{\Sigma}\  
 \mathcal{I}^{\Omega}\mathcal{I}^{\Delta}\mathcal{I}_{\Sigma}
\\
& & \\
& & 
\ +\ \frac{g^{2}}{4|X|^{2}}\mathcal{R}^{\Sigma}
\left[
\mathsf{k}_{\Lambda\, u}\mathsf{k}_{\Sigma}{}^{u}
\ -\ \mathsf{P}_{\Lambda}{}^{x}\mathsf{P}_{\Sigma}{}^{x}\
\right]\; .
\end{array}
\end{equation}

\noindent
Observe that the above equation reduces in the hyperless case, {\em i.e.\/}
$\mathsf{P}_{\Lambda}^{x}=0$, to the expression given in
\cite{Huebscher:2008yz,Meessen:2009ma}.
%%%%%%%%%%%%%%%%%%%%%%%%%%%%%%%%%%%%%%%%%%%%%%%%%%%%%%%%%%%%%%%%%%%%%%
%%%%%%%%%%%%%%%%%%%%%%%%%%%%%%%%%%%%%%%%%%%%%%%%%%%%%%%%%%%%%%%%%%%%%%
%%%%%%%%%%%%%%%%%%%%%%%%%%%%%%%%%%%%%%%%%%%%%%%%%%%%%%%%%%%%%%%%%%%%%%
%%%%%%%%%%%%%%%%%%%%%%%%%%%%%%%%%%%%%%%%%%%%%%%%%%%%%%%%%%%%%%%%%%%%%%
\section{Conclusions}
\label{sec-conclusions}
%%%%%%%%%%%
In this article we have obtained the form of the most general
supersymmetric solution to gauged $N=2$, $d=4$ supergravity coupled to
YM-vector multiplets and hypermultiplets and showed that we are
generically dealing with $\textstyle{1\over 8}$-BPS solutions. The generic form of the
solutions is the one already known from earlier investigations, but
there are some fine differences; for example in ungauged case the
base space is just $\mathbb{R}^{3}$ and in (Abelian) gauged SUGRA
the base space becomes torsionful \cite{Caldarelli:2003pb}, or said differently it must have
a non-trivial $\mathrm{SO}(2)$ holonomy in order to be able to kill
off the effective $\mathrm{U}(1)$ gauge field induced on the
base space.
In our case, see eq.~(\ref{eq:embedding}), we have to face in general
a base space with $\mathrm{SO}(3)$ holonomy as we have to kill off an
effective $\mathrm{SU}(2)$ gauge field.
\par
Clearly, the general equations that need to be solved, such as the
generalized Bogomol'nyi equation in eq.~(\ref{eq:081198}), look daunting
and a general solution is out of reach. But as mentioned in the
introduction, interesting solution can be found and we hope that the
results presented in this article makes finding them easier. An
interesting sub-case to consider would be a theory with an
$\mathrm{SU}(2)$ Fayet-Iliopoulos term along the lines of 
ref.~\cite{Cariglia:2004kk}; work in this direction is in progress.
%%%%%%%%%%%%%%%%%%%%%%%%%%%%%%%%%%%%%%%%%%%%%%%%%%%%%%%%%%%%%%%%%%%%%%
%%%%%%%%%%%%%%%%%%%%%%%%%%%%%%%%%%%%%%%%%%%%%%%%%%%%%%%%%%%%%%%%%%%%%%
%%%%%%%%%%%%%%%%%%%%%%%%%%%%%%%%%%%%%%%%%%%%%%%%%%%%%%%%%%%%%%%%%%%%%%
%%%%%%%%%%%%%%%%%%%%%%%%%%%%%%%%%%%%%%%%%%%%%%%%%%%%%%%%%%%%%%%%%%%%%%
\section*{Acknowledgments}
%%%%%%
The authors would like to thank S.~Vaul\`a and M.~H\"ubscher for their
collaboration in the early stages of this work and D.~Klemm,
K.~Landsteiner, E.~L\'opez, D.~Mansi, C.~Pena and E.~Zorzan for discussions.  This work
has been supported in part by the Spanish Ministry of Science and Education
grant FPA2009-07692, a Ram\'on y Cajal fellowship RYC-2009-05014, the
Comunidad de Madrid grant HEPHACOS S2009ESP-1473, the Princip\'au d'Asturies
grant IB09- 069 and the Spanish Consolider-Ingenio 2010 program CPAN
CSD2007-00042.  TO wishes to thank M.M.~Fern\'andez for her permanent support.
%%%%%%%%%%%%%%%%%%%%%%%%%%%%%%%%%%%%%%%%%%%%%%%%%%%%%%%%%%%%%%%%%%%%%%
%%%%%%%%%%%%%%%%%%%%%%%%%%%%%%%%%%%%%%%%%%%%%%%%%%%%%%%%%%%%%%%%%%%%%%
%%%%%%%%%%%%%%%%%%%%%%%%%%%%%%%%%%%%%%%%%%%%%%%%%%%%%%%%%%%%%%%%%%%%%%
%%%%%%%%%%%%%%%%%%%%%%%%%%%%%%%%%%%%%%%%%%%%%%%%%%%%%%%%%%%%%%%%%%%%%%
\appendix
%%%%%%%%%%%%%%%%%%%%%%%%%%%%%%%%%%%%%%%%%%%%%%%%%%%%%%%%%%%%%%%%%%%%%%
%%%%%%%%%%%%%%%%%%%%%%%%%%%%%%%%%%%%%%%%%%%%%%%%%%%%%%%%%%%%%%%%%%%%%%
%%%%%%%%%%%%%%%%%%%%%%%%%%%%%%%%%%%%%%%%%%%%%%%%%%%%%%%%%%%%%%%%%%%%%%
%%%%%%%%%%%%%%%%%%%%%%%%%%%%%%%%%%%%%%%%%%%%%%%%%%%%%%%%%%%%%%%%%%%%%%

\section{Pauli matrices}
\label{sec-sigmas}

The Hermitean, unitary, traceless, $2\times 2$ Pauli matrices $\sigma^{x}$
($x=1,2,3$) are

\begin{equation}
\label{eq:Paulimatrices}
\sigma^{1} =
\left(
\begin{array}{cr}
0 & 1 \\
1 & 0 \\
\end{array}
\right)\, ,
\hspace{1cm}
\sigma^{2} =
\left(
\begin{array}{cr}
0 & -i \\
i & 0 \\
\end{array}
\right)\, ,
\hspace{1cm}
\sigma^{3} =
\left(
\begin{array}{cr}
1 & 0 \\
0 & -1 \\
\end{array}
\right)\, .
\end{equation}

\noindent
They satisfy the following properties:

\begin{eqnarray}
(\sigma^{x})^{I}{}_{J}(\sigma^{y})^{J}{}_{K} 
& = & 
\delta^{xy}\delta^{I}{}_{K} +i
\varepsilon^{xyz}(\sigma^{z})^{I}{}_{K}\, ,
\\
& & \nonumber \\
\delta^{K}{}_{J}\delta^{L}{}_{I}
& = & 
\tfrac{1}{2}\delta^{K}{}_{I}\delta^{L}{}_{J}
+
\tfrac{1}{2}(\sigma^{m})^{K}{}_{I}(\sigma^{m})^{L}{}_{J}\, ,
\\
& & \nonumber \\
\varepsilon^{IJ}\varepsilon_{KL} 
& = &
-\tfrac{2}{3}(\sigma^{x})^{[I}{}_{[K}(\sigma^{x})^{J]}{}_{L]}\, ,
\\
& & \nonumber \\
(\sigma^{[x|})^{I}{}_{J}(\sigma^{|y]})^{K}{}_{L}
& = & 
-\tfrac{i}{2}\varepsilon^{xyz}[\delta^{I}{}_{L}(\sigma^{z})^{K}{}_{J}
-(\sigma^{z})^{I}{}_{L}\delta^{K}{}_{J}]\, , 
\\
& & \nonumber \\
\varepsilon_{K[I}(\sigma^{x})^{K}{}_{J]} 
& = & 
(\sigma^{x})^{[I}{}_{K}\varepsilon^{J]K} = 0\, ,
\\  
& & \nonumber \\
\varepsilon_{LI}(\sigma^{x})^{I}{}_{J}\varepsilon^{JK} 
& = &
(\sigma^{x})^{K}{}_{L}\, , 
\\
& & \nonumber \\
\left[(\sigma^{x})^{I}{}_{J}\varepsilon^{JK}\right]^{*}
& = & 
-\varepsilon_{IJ}(\sigma^{x})^{J}{}_{K}\, .
\end{eqnarray}

Taking into account the above properties, we have the following decompositions:

\begin{equation}
\label{eq:vectordecomposition}
\begin{array}{rcl}
V^{I}{}_{J} 
&  = & 
{\textstyle\frac{1}{2}}V\ \delta^{I}{}_{J} 
+ {\textstyle\frac{1}{\sqrt{2}}}\ 
V^{x}\, (\sigma^{x})^{I}{}_{J} \, ,  
\\
& & \\
V
& = & 
\delta^{J}{}_{I}V^{I}{}_{J}\, ,    
\\
& & \\
V^{x}{}_{\mu}
&  = & 
 \tfrac{1}{\sqrt{2}}(\sigma^{x})^{J}{}_{I}V^{I}{}_{J}\, ,
\end{array}
\end{equation}

\noindent
where $V$ and $V^{x}$ are real if $V^{I}{}_{J}$ is Hermitian, and

\begin{equation}
\label{eq:2formdecomposition}
\begin{array}{rcl}
A_{IJ} 
&  = & 
{\textstyle\frac{1}{2}}A\ \varepsilon_{IJ} 
+\tfrac{i}{\sqrt{2}} A^{x}\, 
\varepsilon_{IK}(\sigma^{x})^{K}{}_{J}\, ,  
\\
& & \\
A
& = & 
\varepsilon^{IJ}A_{IJ}\, ,    
\\
& & \\
A^{x}
&  = & 
 \tfrac{i}{\sqrt{2}}(\sigma^{x})^{I}{}_{K}\varepsilon^{KJ}A_{IJ}\, ,
\end{array}
\end{equation}

\noindent
or

\begin{equation}
\label{eq:2formdecomposition2}
\begin{array}{rcl}
A^{IJ} 
&  \equiv & (A_{IJ})^{*} 
=
{\textstyle\frac{1}{2}}A^{*}\ \varepsilon^{IJ} 
+\tfrac{i}{\sqrt{2}} A^{x\, *}\, 
(\sigma^{x})^{I}{}_{K}\varepsilon^{KJ}\, ,  
\\
& & \\
A^{*}
& = & 
\varepsilon_{IJ}A^{IJ}\, ,    
\\
& & \\
A^{x\, *}
&  = & 
 \tfrac{i}{\sqrt{2}} \varepsilon_{IK}(\sigma^{x})^{K}{}_{J}A^{IJ}\, .
\end{array}
\end{equation}

In the particular case of the 1- and 2-form spinor bilinears
$\hat{V}^{I}{}_{J}$ and $\hat{\Phi}_{IJ}$ these decompositions are
related by \cite{Bellorin:2005zc}
\begin{equation}
\label{eq:2formdecomposition3}
\hat{\Phi}^{x} = \frac{i}{2X^{*}} 
\left[\hat{V}^{x}\wedge \hat{V} +i\star (\hat{V}^{x}\wedge \hat{V}) \right]\, .  
\end{equation}
%%%%%%%%%%%%%%%%%%%%%%%%%%%%%%%%%%%%%%%%%%%%%%%%%%%%%%%%%%%%%%%%%%%%%%
%%%%%%%%%%%%%%%%%%%%%%%%%%%%%%%%%%%%%%%%%%%%%%%%%%%%%%%%%%%%%%%%%%%%%%
%%%%%%%%%%%%%%%%%%%%%%%%%%%%%%%%%%%%%%%%%%%%%%%%%%%%%%%%%%%%%%%%%%%%%%
%%%%%%%%%%%%%%%%%%%%%%%%%%%%%%%%%%%%%%%%%%%%%%%%%%%%%%%%%%%%%%%%%%%%%%

\section{Gauging holomorphic isometries of Special K\"ahler Geometries}
\label{app-SKGgauging}

In this appendix we will review some basics of the gauging of holomorphic
isometries of the special K\"ahler manifold in $N=2$, $d=4$ supergravities
coupled to vector supermultiplets with the aim of fixing our conventions.

We start by assuming that the Hermitean metric $\mathcal{G}_{ij^{*}}$ admits a
set of Killing vectors\footnote{The index $\Lambda$ always takes values from
  $1$ to $\bar{n}$, but some (or all) the Killing vectors may be zero.}
$\{K_{\Lambda}= k_{\Lambda}{}^{i}\partial_{i}
+k^{*}_{\Lambda}{}^{i^{*}}\partial_{i^{*}}\}$
satisfying the Lie algebra 

\begin{equation}
\label{eq:Liealgebra}
[K_{\Lambda},K_{\Sigma}]= -f_{\Lambda\Sigma}{}^{\Omega} K_{\Omega}\, ,
\end{equation}

\noindent
of the group $G_{V}$ that we want to gauge.

Hermiticity and the $ij$ and $i^{*}j^{*}$ components of the Killing equation
imply that the components $k_{\Lambda}{}^{i}$ and $k^{*}_{\Lambda}{}^{i^{*}}$
of the Killing vectors are, respectively, holomorphic and anti-holomorphic and
satisfy, separately, the above Lie algebra. Once (anti-) holomorphicity is
taken into account, the only non-trivial components of the Killing equation
are

\begin{equation}
\label{eq:Killingeq}
{\textstyle\frac{1}{2}}\pounds_{\Lambda}\mathcal{G}_{ij^{*}}= 
\nabla_{i^{*}}k^{*}_{\Lambda\, j}+ \nabla_{j}k_{\Lambda\, i^{*}} = 0\, ,
\end{equation}

\noindent
where $\pounds_{\Lambda}$ stands for the Lie derivative
w.r.t.~$K_{\Lambda}$. 

The standard $\sigma$-model kinetic term
$\mathcal{G}_{ij^{*}}\partial_{\mu}Z^{i} \partial^{\mu}Z^{*j^{*}}$ is
automatically invariant under infinitesimal reparametrizations of the form

\begin{equation}
\label{eq:deltazio}
\delta_{\alpha} Z^{i} = \alpha^{\Lambda} k_{\Lambda}{}^{i}\, ,  
\end{equation}

\noindent
if the $\alpha^{\Lambda}$s are constants.  If they are arbitrary functions of
the spacetime coordinates $\alpha^{\Lambda}(x)$ we need to introduce a
covariant derivative using as connection the vector fields present in the
theory. The covariant derivative is

\begin{equation}
\label{eq:nablazio}
\mathfrak{D}_{\mu} Z^{i} = \partial_{\mu} Z^{i}+gA^{\Lambda}{}_{\mu} 
k_{\Lambda}{}^{i}\, , 
\end{equation}

\noindent
and transforms as

\begin{equation}
\label{eq:igaugetrans}
\delta_{\alpha}\mathfrak{D}_{\mu} Z^{i}=
\alpha^{\Lambda}(x) \partial_{j} k_{\Lambda}{}^{i}\mathfrak{D}_{\mu} Z^{j}
=-\alpha^{\Lambda}(x)(\pounds_{\Lambda}-K_{\Lambda})
\mathfrak{D}_{\mu} Z^{j} \, ,
\end{equation}

\noindent
provided that the gauge potentials transform as

\begin{equation}
\label{eq:gaugepotentialtransformations}
\delta_{\alpha} A^{\Lambda}{}_{\mu} = 
-g^{-1}\mathfrak{D}_{\mu}\alpha^{\Lambda} 
\equiv
-g^{-1}(\partial_{\mu}\alpha^{\Lambda}
+gf_{\Sigma\Omega}{}^{\Lambda}
A^{\Sigma}{}_{\mu} \alpha^{\Omega})\, .
\end{equation}

For any tensor\footnote{Spacetime and target space tensor indices are not
  explicitly shown.} $\Phi$ transforming covariantly under gauge
transformations, i.e.~tranforming as

\begin{equation}
\delta_{\alpha}\Phi = -\alpha^{\Lambda}(x)(\pounds_{\Lambda}
-K_{\Lambda}) \Phi\, ,  
\end{equation}

\noindent
the gauge covariant derivative is given by 

\begin{equation}
\mathfrak{D}_{\mu}\Phi = 
\{\nabla_{\mu} +\mathfrak{D}_{\mu}Z^{i}\Gamma_{i}  
+\mathfrak{D}_{\mu}Z^{*i^{*}}\Gamma_{i^{*}}  
-gA^{\Lambda}{}_{\mu}(\pounds_{\Lambda} 
-K_{\Lambda})\}\Phi\, .
\end{equation}

\noindent
In particular, on $\mathfrak{D}_{\mu}Z^{i}$

\begin{eqnarray}
\mathfrak{D}_{\mu}  \mathfrak{D}_{\nu} Z^{i} & = & 
\nabla_{\mu}\mathfrak{D}_{\nu} Z^{i}+\Gamma_{jk}{}^{i}\mathfrak{D}_{\mu} Z^{j}
\mathfrak{D}_{\nu} Z^{k} +gA^{\Lambda}{}_{\mu}\partial_{j}k_{\Lambda}{}^{i}
\mathfrak{D}_{\nu} Z^{j}\, ,  \\
& & \nonumber \\
\left[ \mathfrak{D}_{\mu},\mathfrak{D}_{\nu} \right]Z^{i}
& = & gF^{\Lambda}{}_{\mu\nu} k_{\Lambda}{}^{i}\, ,
\end{eqnarray}

\noindent
where 

\begin{equation}
\label{eq:Fdef}
F^{\Lambda}{}_{\mu\nu} = 2\partial_{[\mu}A^{\Lambda}{}_{\nu]}
+gf_{\Sigma\Omega}{}^{\Lambda}A^{\Sigma}{}_{[\mu}A^{\Omega}{}_{\nu]}\, , 
\end{equation}

\noindent
is the gauge field strength and transforms under gauge transformations as

\begin{equation}
\delta_{\alpha} F^{\Lambda}{}_{\mu\nu} = 
-\alpha^{\Sigma}(x)f_{\Sigma\Omega}{}^{\Lambda}\  F^{\Omega}{}_{\mu\nu}\, .
\end{equation}

An important case is that of tensors which only depend on the spacetime
coordinates through the complex scalars $Z^{i}$ and their complex conjugates
so that $\nabla_{\mu} \Phi = \partial_{\mu} \Phi = \partial_{\mu} Z^{i}
\partial_{i} \Phi +\partial_{\mu} Z^{*i^{*}} \partial_{i^{*}} \Phi$. This can
only be true irrespectively of gauge transformations if the tensor $\Phi$ is
invariant, that is

\begin{equation}
\pounds_{\Lambda} \Phi = 0\, .  
\end{equation}

\noindent
The gauge covariant derivative of invariant tensors is always the covariant
pullback of the target covariant derivative:

\begin{equation}
\mathfrak{D}_{\mu}\Phi =
\mathfrak{D}_{\mu}Z^{i}\nabla_{i} \Phi  
+\mathfrak{D}_{\mu}Z^{*i^{*}}\nabla_{i^{*}} \Phi\, .  
\end{equation}

%% The second covariant derivatives are 

%% \begin{equation}
%%   \begin{array}{rcl}
%% \mathfrak{D}_{\mu}  \mathfrak{D}_{\nu}\alpha^{\Lambda} & = & 
%% \nabla_{\mu} \mathfrak{D}_{\nu} \alpha^{\Lambda} 
%% +g f_{\Sigma\Omega}{}^{\Lambda}A^{\Sigma}{}_{\mu}
%% \mathfrak{D}_{\nu}\alpha^{\Lambda}\, ,\\
%% & & \\
%% \left[ \mathfrak{D}_{\mu},\mathfrak{D}_{\nu} \right]\alpha^{\Lambda}
%% & = & gF^{\Sigma}{}_{\mu\nu} f_{\Sigma\Omega}{}^{\Lambda} \alpha^{\Omega}\, ,\\
%% \end{array}
%% \end{equation}

Now, to make the $\sigma$-model kinetic gauge invariant it is enough to
replace the partial derivatives by covariant derivatives.

In $N=2$, $d=4$ supergravity, however, the scalar manifold is not just Hermitean,
but special K\"ahler, and simple isometries of the metric are not necessarily
symmetries of the theory: they must respect the special K\"ahler structure.
Let us first study how the K\"ahler structure is preserved.

The transformations generated by the Killing vectors will preserve the
K\"ahler structure if they leave the K\"ahler potential invariant up to
K\"ahler transformations, i.e., for each Killing vector $K_{\Lambda}$

\begin{equation}
\label{eq:Kconservation}
\pounds_{\Lambda}\mathcal{K}\equiv 
k_{\Lambda}{}^{i}\partial_{i}\mathcal{K}  
+
k^{*}_{\Lambda}{}^{i^{*}}\partial_{i^{*}}\mathcal{K} 
=
\lambda_{\Lambda}(Z)+ \lambda^{*}_{\Lambda}(Z^{*})\, .
\end{equation}

\noindent
{}From this condition it follows that 

\begin{equation}
\label{eq:lambdaalgebra}
%% [\pounds_{\Lambda},\pounds_{\Sigma}]\mathcal{K}= 
%% -f_{\Lambda\Sigma}{}^{\Omega} \pounds_{\Omega} \mathcal{K} =
\pounds_{\Lambda}\lambda_{\Sigma} 
-\pounds_{\Sigma}\lambda_{\Lambda} 
=
-f_{\Lambda\Sigma}{}^{\Omega}\lambda_{\Omega}\, .
\end{equation}

On the other hand, the preservation of the K\"ahler structure implies the
conservation of the K\"ahler 2-form $\mathcal{J}$ 

\begin{equation}
\label{eq:Jconservation}
\pounds_{\Lambda}\mathcal{J}=0\, .
\end{equation}

\noindent
The closedness of $\mathcal{J}$ implies that $\pounds_{\Lambda}\mathcal{J}
= d(i_{k_{\Lambda}}\mathcal{J})$ and therefore the preservation of the
K\"ahler structure implies the existence of a set of real 0-forms
$\mathcal{P}_{\Lambda}$ known as \textit{momentum map} such that

\begin{equation}
\label{eq:SKGmomentummapdef}
i_{k_{\Lambda}}\mathcal{J}= \mathcal{P}_{\Lambda}\, .
\end{equation}

A local solution for this equation is provided by

\begin{equation}
i\mathcal{P}_{\Lambda}
=
k_{\Lambda}{}^{i}\partial_{i}\mathcal{K} -\lambda_{\Lambda}\, ,   
\end{equation}

\noindent
which, on account of eq.~(\ref{eq:Kconservation}) is equivalent to

\begin{equation}
i\mathcal{P}_{\Lambda}
=
-(k^{*}_{\Lambda}{}^{i^{*}}\partial_{i^{*}}\mathcal{K} 
-\lambda^{*}_{\Lambda})\, ,
\end{equation}

\noindent
or

\begin{equation}
\mathcal{P}_{\Lambda}
=
i_{k_{\Lambda}}\mathcal{Q}
-{\textstyle\frac{1}{2i}}(\lambda_{\Lambda}-\lambda^{*}_{\Lambda})\, .
\end{equation}

The momentum map can be used as a prepotential from which the Killing vectors
can be derived:

\begin{equation}
\label{eq:prepo}
k_{\Lambda\, i^{*}} =i\partial_{i^{*}}\mathcal{P}_{\Lambda}\, .  
\end{equation}

Using eqs.~(\ref{eq:Liealgebra}),(\ref{eq:Kconservation}) and
(\ref{eq:lambdaalgebra}) one finds

\begin{equation}
\label{eq:momentummaptransformationrule}
\pounds_{\Lambda}\mathcal{P}_{\Sigma} = 
2i k_{[\Lambda}{}^{i}k^{*}_{\Sigma]}{}^{j^{*}}\mathcal{G}_{ij^{*}} =
-f_{\Lambda\Sigma}{}^{\Omega} \mathcal{P}_{\Omega}\, .  
\end{equation}

The gauge transformation rule a symplectic section $\Phi$ of K\"ahler weight
$(p,q)$ is\footnote{Again, spacetime and target space tensor indices are not
  explicitly shown. Symplectic indices are not shown, either.}

\begin{equation}
\delta_{\alpha}\Phi = -\alpha^{\Lambda}(x)(\mathbb{L}_{\Lambda}
-K_{\Lambda}) \Phi\, ,  
\end{equation}

\noindent
where $\mathbb{L}_{\Lambda}$ stands for the symplectic and K\"ahler-covariant
Lie derivative w.r.t.~$K_{\Lambda}$ and is given by

\begin{equation}
\mathbb{L}_{\Lambda} \Phi = \{\pounds_{\Lambda}
-[\mathcal{S}_{\Lambda}
-{\textstyle\frac{1}{2}}(p\lambda_{\Lambda}+q\lambda^{*}_{\Lambda})]\}
\Phi\, ,  
\end{equation}

\noindent
where the $\mathcal{S}_{\Lambda}$ are $\mathfrak{sp}(\bar{n},\mathbb{R})$ matrices that
provide a representation of the Lie algebra of the gauge group $G_{V}$:
\begin{equation}
\label{eq:SLiealgebra}
[\mathcal{S}_{\Lambda},\mathcal{S}_{\Sigma}]= 
+f_{\Lambda\Sigma}{}^{\Omega} \mathcal{S}_{\Omega}\, .
\end{equation}

\noindent
The gauge covariant derivative acting on these sections is given by 

\begin{equation}
  \begin{array}{rcl}
\mathfrak{D}_{\mu}\Phi & = &  
\{\nabla_{\mu} +\mathfrak{D}_{\mu}Z^{i}\Gamma_{i}  
+\mathfrak{D}_{\mu}Z^{*i^{*}}\Gamma_{i^{*}}  
+{\textstyle\frac{1}{2}}(pk_{\Lambda}{}^{i}\partial_{i}\mathcal{K}+
qk^{*}_{\Lambda}{}^{i^{*}}\partial_{i^{*}}\mathcal{K})  \\
& & \\
& & 
+gA^{\Lambda}{}_{\mu}[\mathcal{S}_{\Lambda} 
+{\textstyle\frac{i}{2}}(p-q)\mathcal{P}_{\Lambda} -(\pounds_{\Lambda} 
-K_{\Lambda})]\}\Phi\, .\\
\end{array}
\end{equation}

Invariant sections are those for which 

\begin{equation}
\mathbb{L}_{\Lambda} \Phi =0\, ,\,\,\,\, \Rightarrow\,\,\,\,  
\pounds_{\Lambda}\Phi = 
[\mathcal{S}_{\Lambda}
-{\textstyle\frac{1}{2}}(p\lambda_{\Lambda}+q\lambda^{*}_{\Lambda})]
\Phi\, ,
\end{equation}

\noindent
and their gauge covariant derivatives are, again, the covariant pullbacks of
the K\"ahler-covariant derivatives:

\begin{equation}
\mathfrak{D}_{\mu}\Phi = \mathfrak{D}_{\mu}Z^{i}\mathcal{D}_{i}\Phi  
+\mathfrak{D}_{\mu}Z^{*i^{*}}\mathcal{D}_{i^{*}}\Phi\, .  
\end{equation}

By hypothesis (preservation of the special K\"ahler structure), the canonical
weight $(1,-1)$ section $\mathcal{V}$ is an invariant section

\begin{equation}
\label{eq:KLV}
K_{\Lambda}\mathcal{V} = 
[\mathcal{S}_{\Lambda}
-{\textstyle\frac{1}{2}}(\lambda_{\Lambda}-\lambda^{*}_{\Lambda})]
\mathcal{V}\, ,
%% \hspace{1cm}
%% \mathcal{S}_{\Lambda}\mathcal{L}^{\Sigma}=
%% f_{\Lambda\Omega}{}^{\Sigma}\mathcal{L}^{\Omega}\, ,
%% \hspace{1cm}
%% \mathcal{S}_{\Lambda}\mathcal{M}_{\Sigma}=
%% f_{\Lambda\Sigma}{}^{\Omega}\mathcal{M}_{\Omega}\, ,
\end{equation}

\noindent
and its gauge covariant derivative is given by

\begin{equation}
\mathfrak{D}_{\mu}\mathcal{V} = 
\mathfrak{D}_{\mu}Z^{i}\mathcal{D}_{i}\mathcal{V}= 
\mathfrak{D}_{\mu}Z^{i}\mathcal{U}_{i}\, .
\end{equation}

\noindent
Using the covariant holomorphicity of $\mathcal{V}$ one can write 

\begin{equation}
K_{\Lambda}\mathcal{V} = k_{\Lambda}{}^{i}\mathcal{U}_{i}
-i\mathcal{P}_{\Lambda}\mathcal{V} -{\textstyle\frac{1}{2}}
(\lambda_{\Lambda}-\lambda^{*}_{\Lambda})\mathcal{V}\, ,
\end{equation}

\noindent
and, comparing with eq.~(\ref{eq:KLV}) and taking the symplectic product with
$\mathcal{V}^{*}$, we find another expression for the momentum map

\begin{equation}
\label{eq:altmomentum}
\mathcal{P}_{\Lambda} = \langle\, \mathcal{V}^{*} \mid
\mathcal{S}_{\Lambda}\mathcal{V}\, \rangle\, ,  
\end{equation}

\noindent
which leads, via eq.~(\ref{eq:prepo}) to another expression for the Killing
vectors

\begin{equation}
\label{eq:altkilling}
k_{\Lambda}{}^{i}=  i\partial^{i}\mathcal{P}_{\Lambda}=
i\langle\, \mathcal{V} \mid
\mathcal{S}_{\Lambda}\mathcal{U}^{*i}\, \rangle\, .
\end{equation}

\noindent
If we take the symplectic product with $\mathcal{V}$ instead, we get the
following condition 

\begin{equation}
\label{eq:VSLV}
\langle\, \mathcal{V} \mid
\mathcal{S}_{\Lambda}\mathcal{V}\, \rangle  =0\, .
\end{equation}

\noindent
Using the same identity and $\mathcal{G}_{ij^{*}}=
-i \langle\, \mathcal{U}_{i} \mid \mathcal{U}^{*}_{j^{*}}\, \rangle$ 
one can also show that

\begin{equation}
\label{eq:klks}
k_{\Lambda}{}^{i}k^{*}_{\Sigma}{}^{j^{*}}\mathcal{G}_{ij^{*}}=
\mathcal{P}_{\Lambda}  \mathcal{P}_{\Sigma} -i
\langle\, \mathcal{S}_{\Lambda}\mathcal{V}\mid
\mathcal{S}_{\Sigma}\mathcal{V}^{*}\, \rangle\, .
\end{equation}

It follows that
\begin{equation}
\langle\, \mathcal{S}_{[\Lambda}\mathcal{V} \mid
\mathcal{S}_{\Sigma]}\mathcal{V}^{*}\, \rangle\,=
-{\textstyle\frac{1}{2}}f_{\Lambda\Sigma}{}^\Omega\mathcal P_\Omega.
\end{equation}

The gauge covariant derivative of $\mathcal{U}_{i}$ is

\begin{equation}
\mathfrak{D}_{\mu}\mathcal{U}_{i} = 
\mathfrak{D}_{\mu}Z^{j}\mathcal{D}_{j}\mathcal{U}_{i} 
+\mathfrak{D}_{\mu}Z^{*j^{*}}\mathcal{D}_{j^{*}}\mathcal{U}_{i}
= 
i\mathcal{C}_{ijk}\mathcal{U}^{*j}
\mathfrak{D}_{\mu}Z^{k}
+\mathcal{G}_{ij^{*}}\mathcal{V}\mathfrak{D}_{\mu}Z^{*j^{*}}\, .  
\end{equation}

Then, explicitly, the covariant derivative on the upper
$\mathcal{L}^{\Lambda}$ and lower $\mathcal{M}_{\Lambda}$ components of the
canonical section and on the supersymmetry parameters $\epsilon_{I}$, which
are of $(\textstyle{1\over 2},-\textstyle{1\over 2})$ weight, are given by 

\begin{eqnarray}
\label{eq:dLL}
\mathfrak{D}_{\mu} \mathcal{L}^{\Lambda}
& = & 
\partial_{\mu} \mathcal{L}^{\Lambda}
+i\hat{\mathcal{Q}}_{\mu} \mathcal{L}^{\Lambda}
+gA^{\Omega}f_{\Omega\Sigma}{}^{\Lambda} \mathcal{L}^{\Sigma}\, ,
\\
& & \nonumber \\
\label{eq:dML}
\mathfrak{D}_{\mu} \mathcal{M}_{\Lambda}
& = & 
\partial_{\mu} \mathcal{M}_{\Lambda}
+i\hat{\mathcal{Q}}_{\mu} \mathcal{M}_{\Lambda}
-gA^{\Omega}f_{\Omega\Lambda}{}^{\Sigma} \mathcal{M}_{\Sigma}\, ,
\\
& & \nonumber \\
\label{eq:dei}
\mathfrak{D}_{\mu}\epsilon_{I} 
& = &
\left\{
\nabla_{\mu}+{\textstyle\frac{i}{2}}\hat{\mathcal{Q}}_{\mu}
\right\}\epsilon_{I}\, ,
\end{eqnarray}

\noindent
where we have defined

\begin{equation}
\hat{\mathcal{Q}}_{\mu}
\equiv \mathcal{Q}_{\mu}+gA^{\Lambda}{}_{\mu}\mathcal{P}_{\Lambda}\, .  
\end{equation}

The formalism, so far, applies to any group $G_{V}$ of isometries. However, we
will restrict ourselves to those for which the matrices 

\begin{equation}
\mathcal{S}_{\Lambda} =
\left(
  \begin{array}{cc}
a_{\Lambda}{}^{\Omega}{}_{\Sigma} & b_{\Lambda}{}^{\Omega\Sigma} \\
& \\
c_{\Lambda\Omega\Sigma} & d_{\Lambda\Omega}{}^{\Sigma} \\
\end{array}
\right)\, , 
\end{equation}

\noindent
have $b=c=0$. The symplectic transformations with $b\neq 0$ are not symmetries
of the action and the gauging of symmetries with $c\neq 0$ leads to the
presence of complicated Chern-Simons terms in the action. The matrices $a$ and
$d$ are 

\begin{equation}\label{eq:GGchoice}
a_{\Lambda}{}^{\Omega}{}_{\Sigma} =f_{\Lambda\Sigma}{}^{\Omega}\, ,
\hspace{1cm}
d_{\Lambda\Omega}{}^{\Sigma} =  -f_{\Lambda\Omega}{}^{\Sigma}\, .
\end{equation}

\noindent
These restrictions lead to additional identities. First, observe that the
condition eq.~(\ref{eq:VSLV}) takes the form

\begin{equation}
\label{eq:VSLV2}
f_{\Lambda\Sigma}{}^{\Omega}\mathcal{L}^{\Sigma}\mathcal{M}_{\Omega}=0\, ,
\end{equation}

\noindent
and the covariant derivative of eq.~(\ref{eq:VSLV}) $\langle\, \mathcal{V}
\mid \mathcal{S}_{\Lambda}\mathcal{U}_{i}\, \rangle =0$

\begin{equation}
\label{eq:VSLV3}
f_{\Lambda\Sigma}{}^{\Omega}(f^{\Sigma}{}_{i}\mathcal{M}_{\Omega}
+h_{\Omega\, i}\mathcal{L}^{\Sigma})=0\, .
\end{equation}

\noindent
Then, using eqs.~(\ref{eq:altmomentum}) and (\ref{eq:altkilling}) and
eqs.~(\ref{eq:VSLV}),(\ref{eq:VSLV2})  and (\ref{eq:VSLV3}) we find that

\begin{eqnarray}
\label{eq:LP0}
\mathcal{L}^{\Lambda}\mathcal{P}_{\Lambda} & = & 0\, ,\\
& & \nonumber \\
\label{eq:LK0}
\mathcal{L}^{\Lambda}k_{\Lambda}{}^{i} & = & 0\, ,\\  
& & \nonumber \\
\label{eq:LKfP}
\mathcal{L}^{*\Lambda}k_{\Lambda}{}^{i} & = & 
-if^{*\Lambda\, i}\mathcal{P}_{\Lambda}\, .  
\end{eqnarray}

{}From the first two equations it follows that

\begin{equation}
\mathcal{L}^{\Lambda}\lambda_{\Lambda}=0\, .  
\end{equation}

Some further equations that can be derived and are extensively used in the
calculation throughout the text are explicit versions of
Eqs. (\ref{eq:altmomentum}) and (\ref{eq:altkilling}), {\em i.e.\/}
\begin{equation}
\label{eq:1}
\mathcal{P}_{\Lambda} \; =\; 2f_{\Lambda\Sigma}{}^{\Gamma}
\Re{\rm e}\,\left( \mathcal{L}^{\Sigma} \mathcal{M}^{*}_{\Gamma}\right)\, ,
\hspace{1cm}
k_{\Lambda\, i^{*}} \; =\; 
if_{\Lambda\Sigma}{}^{\Gamma}\left( f^{*\Sigma}_{i^{*}}M_{\Gamma} 
+ \mathcal{L}^{\Sigma}h^{*}_{\Gamma i^{*}}
 \right) \; .
\end{equation}

Finally, notice the identity

\begin{equation}
\label{eq:laleche}
k_{\Lambda\, i^{*}}\mathfrak{D}Z^{*i^{*}}  
-k^{*}_{\Lambda i}\mathfrak{D}Z^{i} =i\mathfrak{D}\mathcal{P}_{\Lambda}
= i(d\mathcal{P}_{\Lambda}
+gf_{\Lambda\Sigma}{}^{\Omega}A^{\Sigma}\mathcal{P}_{\Omega}) \, . 
\end{equation}

The absolutely last comment in this appendix is the following: if we start
from the existence of a prepotential $\mathcal{F}(\mathcal{X})$, then
eq.~(\ref{eq:VSLV}) implies

\begin{equation}
\label{eq:2PreP}
0 \; =\; f_{\Lambda\Sigma}{}^{\Gamma}\ \mathcal{X}^{\Sigma}\partial_{\Gamma}\ 
\mathcal{F} \; , 
\end{equation}

\noindent
the meaning of which is that one can gauge only the invariances of the
prepotential.  To put it differently: if you want to construct a model having
$\mathfrak{g}$ as the gauge algebra, you need to pick a prepotential that is
$\mathfrak{g}$-invariant.
%%%%%%%%%%%%%%%%%%%%%%%%%%%%%%%%%%%%%%%%%%%%%%%%%%%%%%%%%%%%%%%%%%%%%%
%%%%%%%%%%%%%%%%%%%%%%%%%%%%%%%%%%%%%%%%%%%%%%%%%%%%%%%%%%%%%%%%%%%%%%
%%%%%%%%%%%%%%%%%%%%%%%%%%%%%%%%%%%%%%%%%%%%%%%%%%%%%%%%%%%%%%%%%%%%%%
%%%%%%%%%%%%%%%%%%%%%%%%%%%%%%%%%%%%%%%%%%%%%%%%%%%%%%%%%%%%%%%%%%%%%%
\section{Gauging  isometries of quaternionic K\"ahler manifolds}
\label{app-QMgauging}
%%%%
We start by assuming that the metric $\mathsf{H}_{uv}$ admits Killing vectors
$\mathsf{k}_{\Lambda}{}^{u}$ satisfying the Lie algebra

\begin{equation}
\label{eq:Liealgebraquat}
[k_{\Lambda},k_{\Sigma}]= -f_{\Lambda\Sigma}{}^{\Omega} k_{\Omega}\, ,
\end{equation}

\noindent
where, as in previous cases, for certain values of $\Lambda$ the vectors and
the structure constants can vanish. The metric and the ungauged sigma model
are invariant under the global transformations

\begin{equation}
\label{eq:globalqisometries}
\delta_{\alpha}q^{u} = \alpha^{\Lambda}\mathsf{k}_{\Lambda}{}^{u}(q)\, .  
\end{equation}

In order to make this global invariance local, we just have to replace the
standard derivatives of the scalars by the covariant derivatives

\begin{equation}
\label{eq:Dq}
\mathfrak{D}_{\mu}q^{u} \equiv \partial_{\mu}q^{u} +gA^{\Lambda}{}_{\mu}
\mathsf{k}_{\Lambda}{}^{u}\, ,  
\end{equation}

\noindent
which will transform according to

\begin{equation}
\delta_{\alpha} \mathfrak{D}_{\mu}q^{u}
=
\alpha^{\Lambda}(x)\partial_{v}\mathsf{k}_{\Lambda}{}^{u}  
 \mathfrak{D}_{\mu}q^{v}\, ,
\end{equation}

\noindent
provided that the gauge potentials transform in the standard form
eq.~(\ref{eq:gaugepotentialtransformations}).

This is enough to gauge the global symmetry of the scalars' kinetic term.
However, the isometries of the metric need not be global symmetries of the
full supergravity theory. They have to preserve the quaternionic-K\"ahler
structure as well, and not just the metric. In order to discuss the
preservation of this structure, we need to define $\mathrm{SU}(2)$-covariant Lie
derivatives.

Let $\psi^{x}(q)$ be a field on $\mathsf{HM}$ transforming under infinitesimal
local $\mathrm{SU}(2)$ transformations according to 

\begin{equation}
\delta_{\lambda} \psi^{x} = -\varepsilon^{xyz}\lambda^{y}\psi^{z}\, .
\end{equation}

\noindent
Its $\mathrm{SU}(2)$-covariant derivative is given by 

\begin{equation}
\mathsf{D} \psi^{x} 
=
d\psi^{x}+\varepsilon^{xyz}\mathsf{A}^{y}\psi^{z}\, ,  
\end{equation}

\noindent
where the $\mathrm{SU}(2)$-connection 1-form transforms as

\begin{equation}
\delta_{\lambda} \mathsf{A}^{x} = \mathsf{D} \lambda^{x}\, .
\end{equation}

To define an $\mathrm{SU}(2)$-covariant Lie derivative with respect to the Killing
vector $\mathsf{k}_{\Lambda}$ $\mathbb{L}_{\Lambda}$, we add to the standard
one $\pounds_{\Lambda}$ a local $\mathrm{SU}(2)$ transformation whose transformation
parameter is given by the compensator field $\mathsf{W}_{\Lambda}{}^{x}$:

\begin{equation}
\mathbb{L}_{\Lambda}\psi^{x} \equiv \pounds_{\Lambda}\psi^{x}
+\varepsilon^{xyz}\mathsf{W}_{\Lambda}{}^{y}\psi^{z}\, ,  
\end{equation}

\noindent
which is such that

\begin{equation}
\delta_{\lambda} \mathsf{W}_{\Lambda}{}^{x}
= \pounds_{\Lambda}\lambda^{x}
-\varepsilon^{xyz}\lambda^{y}\mathsf{W}_{\Lambda}{}^{z}
=\mathbb{L}_{\Lambda}\lambda^{x}\, .
\end{equation}

$\mathbb{L}_\Lambda$ is clearly a linear operator which satisfies the Leibnitz
rule for scalar and vector products of ${\rm SU}(2)$ vectors. The Lie
derivative must also satisfy

\begin{equation}
\left[\mathbb{L}_{\Lambda},\,\mathbb{L}_{\Sigma}\right]
=
\mathbb{L}_{[\mathsf{k}_{\Lambda},\,\mathsf{k}_{\Sigma}]}
\end{equation}

\noindent
which implies the Jacobi identity. This requires 

\begin{equation}
\label{eq:Wcurl}
\pounds_{\Lambda}\mathsf{W}_{\Sigma}{}^{x}-\pounds_{\Sigma}\mathsf{W}_{\Lambda}{}^{x}
+\varepsilon^{xyz}\mathsf{W}_{\Lambda}{}^{y}\mathsf{W}_{\Sigma}^{z}=
-f_{\Lambda\Sigma}{}^{\Gamma}\mathsf{W}_{\Gamma}{}^{x}\, ,
\end{equation}

\noindent
where, due to the assumed linear dependency of $\mathsf{W}_{\Lambda}$ on
$\mathsf{k}_{\Lambda}$,
$\mathsf{W}_{[\mathsf{k}_{\Lambda},\,\mathsf{k}_{\Sigma}]}
=-f_{\Lambda\Sigma}{}^{\Gamma}\mathsf{W}_{\Gamma}$.

In order to satisfy equation (\ref{eq:Wcurl}) we introduce another ${\rm
  SU}(2)$ vector $\mathsf{P}_{\Lambda}{}^{x}$ such that 

\begin{equation}
\label{eq:triholomomentummapdef}
\mathsf{W}_{\Lambda}{}^{x}
\equiv
\mathsf{k}_{\Lambda}{}^{u}\mathsf{A}^{x}{}_{u}-\mathsf{P}_{\Lambda}{}^{x}\, ,
\end{equation}

\noindent
which  has to satisfy the equivariance condition

\begin{equation}
\mathsf{D}_{\Lambda}\mathsf{P}_{\Sigma}{}^{x}
-\mathsf{D}_{\Sigma}\mathsf{P}_{\Lambda}{}^{x}
-\varepsilon^{xyz} \mathsf{P}_{\Lambda}{}^{y}\mathsf{P}_{\Sigma}{}^{z}
-\varkappa\ \mathsf{k}_{\Lambda}{}^{u}\mathsf{k}_{\Sigma}{}^{v}\,
\mathsf{K}^{x}{}_{uv}
=
-f_{\Lambda\Sigma}{}^{\Gamma}\mathsf{P}_{\Gamma}{}^{x}\, ,
\label{eq:Pcurl}
\end{equation} 

\noindent
where $\mathsf{D}_{\Lambda}\equiv \mathsf{k}_{\Lambda}{}^{u}\mathsf{D}_{u}$
and we have used the fundamental property of the hyperK\"ahler manifolds

\begin{equation}
\label{eq:FvsK}
\mathsf{F}^{x}\; =\;  \varkappa\ \mathsf{K}^{x}\, ,
\end{equation}

\noindent
where 

\begin{equation}
\mathsf{F}^{x}\; \equiv\; d\mathsf{A}^{x} 
\ +\ {\textstyle\frac{1}{2}}\varepsilon^{xyz}\
     \mathsf{A}^{y} \wedge \mathsf{A}^{z}\, , 
\end{equation}

\noindent
is the field strength of the $\mathrm{SU}(2)$-connection and $\varkappa$ is a
non-vanishing real number which has to be negative for the kinetic energy of
the hyperscalars to be positive; we take $\varkappa = -2$
as to have a conventionally defined kinetic term for the hyperscalars.

$\mathsf{P}_{\Lambda}{}^{x}$ is going to
be the \textit{triholomorphic momentum map} when we impose the preservation of
the hyperK\"ahler structure $\mathsf{K}^{x}$ by the global transformations
eq.~(\ref{eq:globalqisometries}) and the compensating $\mathrm{SU}(2)$ transformation
with parameter $\mathsf{W}_{\Lambda}$. This condition is expressed using
$\mathbb{L}$ as

\begin{equation}
\mathbb{L}_{\Lambda}\mathsf{K}^{x}{}_{uv} 
= 
\pounds_{\Lambda}\mathsf{K}^{x}{}_{uv}
+\varepsilon^{xyz}(\mathsf{k}_{\Lambda}{}^{w}\mathsf{A}^{y}{}_{w} 
-\mathsf{P}_{\Lambda}{}^{y})\mathsf{K}^{z}{}_{uv}
=
-2\mathsf{D}_{[u|}(\mathsf{k}_{\Lambda}{}^{w}\mathsf{K}^{x}{}_{w|v]}) 
-\varepsilon^{xyz}\mathsf{P}_{\Lambda}{}^{y}\mathsf{K}^{z}{}_{uv}
=0\, .  
\end{equation}

\noindent
Using the covariant constancy of the hyperK\"ahler structure, this condition
can be rewritten in the form

\begin{equation}
 2(\nabla_{[u|}\mathsf{k}_{\Lambda}{}^{w})\mathsf{K}^{x}{}_{w|v]}
-\varepsilon^{xyz} \mathsf{P}_{\Lambda}{}^{y}\mathsf{K}^{z}{}_{uv}=0\, ,
\end{equation}

\noindent
and, contracting the whole equation with $\mathsf{K}^{y\, uv}$ we find

\begin{equation}
\mathsf{K}^{x\, uv}\nabla_{u}\mathsf{k}_{\Lambda\, v} =
  -2m\mathsf{P}_{\Lambda}{}^{x}\, .
\end{equation}

\noindent
Acting on both sides of this equations with $\mathsf{D}_{w}$ and using the
Killing vector identity $\nabla_{w}\nabla_{u}\mathsf{k}_{\Lambda\, v}= 
R_{wruv}\mathsf{k}_{\Lambda}{}^{r}$ we get 

\begin{equation}
\mathsf{k}_{\Lambda}{}^{r}R_{wruv}\mathsf{K}^{x\, uv} =
-2m\mathsf{D}_{w}\mathsf{P}_{\Lambda}{}^{x}\, .   
\end{equation}

\noindent
Finally, using 

\begin{equation}
\label{eq:KxUU}
\mathsf{K}^{x}{}_{uv} =
-i \sigma^{x}{}_{IJ}\mathsf{U}^{\alpha I}{}_{u}\mathsf{U}^{\beta J}{}_{v}
\mathbb{C}_{\alpha\beta}\, ,
\hspace{1cm}
\sigma^{x}{}_{IJ}\equiv\sigma^{x}{}_{I}{}^{K} \varepsilon_{JK}\, ,  
\end{equation}

\noindent 
in  

\begin{equation}
\label{eq:relationcurvatures}
R_{ts}{}^{uv}\ \mathsf{U}^{\alpha I}{}_{u}\ \mathsf{U}^{\beta J}{}_{v} \; =\;
        -\mathsf{G}_{ts}^{IJ}\ \mathbb{C}^{\alpha\beta}
        \ -\ \overline{\mathsf{R}}_{ts}^{\ \alpha\beta}\ \varepsilon^{IJ}
    \; =\;  \mathsf{F}_{ts}^{IJ}\ \mathbb{C}^{\alpha\beta}
        \ -\ \overline{\mathsf{R}}_{ts}^{\ \alpha\beta}\ \varepsilon^{IJ} \; ,
\end{equation}

\noindent
we get

\begin{equation}
R_{wruv}\mathsf{K}^{x\, uv} 
=
-2m\ \mathsf{F}^{x}{}_{wr}
=
-2m\varkappa\ \mathsf{K}^{x}{}_{wr}\, . 
\end{equation}

\noindent
Substituting above, we arrive at 

\begin{equation}
\label{eq:DPxKk}
\mathsf{D}_{u}\mathsf{P}_{\Lambda}{}^{x} 
=
\varkappa\ \mathsf{K}^{x}{}_{uv} \mathsf{k}_{\Lambda}{}^{v}\, ,
\end{equation}

\noindent
which can be taken as the defining equation of the triholomorphic momentum
map. From this equation we find

\begin{equation}
\label{eq:DPkK}
\mathsf{D}_{\Sigma}\mathsf{P}_{\Lambda}{}^{x} 
=
\varkappa\ \mathsf{k}_{\Sigma}{}^{u}\mathsf{k}_{\Lambda}{}^{v}
\mathsf{K}^{x}{}_{uv}\, ,
\end{equation}

\noindent
and, substituting directly in eq.~(\ref{eq:Pcurl}) we obtain

\begin{equation}
\label{eq:LPx0}
\mathbb{L}_{\Lambda}\mathsf{P}_{\Sigma}{}^{x}
=\mathsf{D}_{\Lambda}\mathsf{P}_{\Sigma}{}^{x}
-\varepsilon^{xyz}\mathsf{P}_{\Lambda}{}^{y}\mathsf{P}_{\Sigma}{}^{z}  
+f_{\Lambda\Sigma}{}^{\Omega}\mathsf{P}_{\Omega}{}^{x} =0\, ,
\end{equation}

\noindent
which says that the triholomorphic momentum map is an invariant field and 

\begin{equation}
\varepsilon^{xyz}\mathsf{P}_{\Lambda}{}^{y}\mathsf{P}_{\Sigma}{}^{z}  
-\varkappa\  
\mathsf{k}_{\Lambda}{}^{u}\mathsf{k}_{\Sigma}{}^{v}\mathsf{K}^{x}{}_{uv}
=
f_{\Lambda\Sigma}{}^{\Omega}\mathsf{P}_{\Omega}{}^{x}\, .
\end{equation}

Now, for a field $\Phi$ (possibly with spacetime, quaternionic, $\mathrm{SU}(2)$ or
gauge indices) which under eq.~(\ref{eq:globalqisometries}) transforms according
to

\begin{equation}
\delta_{\alpha}\Phi =-\alpha (\mathbb{L}_{\Lambda}-\mathsf{k}_{\Lambda})\Phi\, ,  
\end{equation}

\noindent
we define the gauge covariant derivative

\begin{equation}
\mathfrak{D}_{\mu}\Phi
\equiv
\{\nabla_{\mu} +\mathfrak{D}_{\mu}q^{u}\Gamma_{u} 
-gA^{\Lambda}{}_{\mu}(\mathbb{L}_{\Lambda}-\mathsf{k}_{\Lambda})
+\mathfrak{D}_{\mu}q^{u}\mathsf{A}^{x}{}_{u}
\}\Phi  \, .
\end{equation}

For the triholomorphic momentum map, we have, on account of 
eq.~(\ref{eq:LPx0}), which we can rewrite in the form

\begin{equation}
k_{\Lambda}{}^{u}\partial_{u}\mathsf{P}_{\Sigma}{}^{x} =
- \varepsilon^{xyz}(k_{\Lambda}{}^{u}\mathsf{A}^{y}{}_{u}
-\mathsf{P}_{\Lambda}{}^{y})\mathsf{P}_{\Sigma}{}^{z} 
-f_{\Lambda\Sigma}{}^{\Omega} \mathsf{P}_{\Omega}{}^{x}\, ,
\end{equation}
 
\noindent
the following expressions for its  gauge covariant derivative

\begin{eqnarray}
\mathfrak{D}_{\mu}\mathsf{P}_{\Lambda}{}^{x}
& = & 
\partial_{\mu}  \mathsf{P}_{\Lambda}{}^{x}
+\varepsilon^{xyz}\hat{\mathsf{A}}^{y}{}_{\mu} \mathsf{P}_{\Lambda}{}^{z} 
+f_{\Lambda\Sigma}{}^{\Omega}A^{\Sigma}{}_{\mu} \mathsf{P}_{\Omega}{}^{x}\,
,\\
& & \nonumber \\
\label{eq:DPxDqDPx}
\mathfrak{D}_{\mu}\mathsf{P}_{\Lambda}{}^{x}
& = & 
\mathfrak{D}_{\mu}q^{u}
\mathsf{D}_{u}\mathsf{P}_{\Lambda}{}^{x}\, ,
\end{eqnarray}

\noindent
where we have defined 

\begin{equation}
\hat{\mathsf{A}}^{x}{}_{\mu}
\equiv 
\partial_{\mu}q^{u}\mathsf{A}^{x}{}_{u} 
+gA^{\Lambda}{}_{\mu}\mathsf{P}_{\Lambda}{}^{x}\, . 
\end{equation}

Under eq.~(\ref{eq:globalqisometries}), spinors with $\mathrm{SU}(2)$ indices undergo
the following transformation

\begin{equation}
\delta_{\alpha}\psi_{I} =-\alpha^{\Lambda}\mathsf{W}_{\Lambda}{}^{x}
{\textstyle\frac{i}{2}} \sigma^{x}{}_{I}{}^{J}\psi_{J}\, .  
\end{equation}

\noindent
Then, using the general formula, their covariant derivative is given by 

\begin{equation}
\mathfrak{D}_{\mu}  \psi_{I}
=
\nabla_{\mu}\psi_{I} 
+\hat{\mathsf{A}}^{x}{}_{\mu}\, {\textstyle\frac{i}{2}}
\sigma^{x}{}_{I}{}^{J}\psi_{J}\, .
\end{equation}

If we take into account their K\"ahler weight and possible gaugings of the
isometries of the special-K\"ahler manifold, we have for the supersymmetry
parameters of $N=2$, $d=4$ supergravity

\begin{equation}
\label{eq:covariantderivativeonepsilon}
\mathfrak{D}_{\mu}  \epsilon_{I}
=
\{\nabla_{\mu}+{\textstyle\frac{i}{2}}\hat{\mathcal{Q}}_{\mu}\} \epsilon_{I} 
+\hat{\mathsf{A}}^{x}{}_{\mu}{\textstyle\frac{i}{2}}
\sigma^{x}{}_{I}{}^{J}\epsilon_{J}\, .
\end{equation}

%%%%%%%%%%%%%%%%%%%%%%%%%%%%%%%%%%%%%%%%%%%%%%%%%%%%%%%%%%%%%%%%%%%%%%
%%%%%%%%%%%%%%%%%%%%%%%%%%%%%%%%%%%%%%%%%%%%%%%%%%%%%%%%%%%%%%%%%%%%%%
%%%%%%%%%%%%%%%%%%%%%%%%%%%%%%%%%%%%%%%%%%%%%%%%%%%%%%%%%%%%%%%%%%%%%%
%%%%%%%%%%%%%%%%%%%%%%%%%%%%%%%%%%%%%%%%%%%%%%%%%%%%%%%%%%%%%%%%%%%%%%
%%%%%%%%%%%%%%%%%%%%%%%%%%%%%%%%%%%%%%%%%%%%%%%%%%%%%%%%%%%%%%%%%%%%%%
%%%%%%%%%%%%%%%%%%%%%%%%%%%%%%%%%%%%%%%%%%%%%%%%%%%%%%%%%%%%%%%%%%%%%%

\end{document}